\documentclass{aa}  
\usepackage{comment}
\usepackage{siunitx}
\usepackage{amsmath}
\usepackage{subcaption}
\usepackage{nicefrac}
\usepackage{graphicx}
\usepackage{color}
\usepackage{hyperref}
\usepackage{csvsimple}
\usepackage{ulem}
\usepackage[switch]{lineno}
\usepackage{booktabs,wrapfig}
\hypersetup{colorlinks,linkcolor={blue},citecolor={blue}}

\usepackage{textcomp,gensymb}
\usepackage{caption} 
\usepackage{tabularx}
\usepackage[bottom]{footmisc}
\usepackage[varg]{txfonts}

\usepackage{listings}
\usepackage{xcolor}

\lstdefinelanguage{ADQL}{
  morekeywords={
    SELECT,FROM,WHERE,AND,OR,NOT,IN,AS,JOIN,ON,ORDER,BY,GROUP,HAVING,LIMIT,TOP,
    DISTINCT,COUNT,SUM,AVG,MIN,MAX
  },
  sensitive=false,
  morecomment=[l]--,
  morestring=[b]',
}

\begin{document} 

    \title{Dormant black hole candidates from \textit{Gaia} DR3 summary diagnostics}

   \author{Johanna M\"uller-Horn\inst{1,2}\thanks{Corresponding author \email{mueller-horn@mpia.de}} \and
          Hans-Walter Rix\inst{1} \and
          Kareem El-Badry\inst{3,1} \and
          Ben Pennell\inst{1,2} \and
          Matthew Green\inst{1,4,5} \and
          Jiadong Li\inst{1} \and
          Rhys Seeburger\inst{1}}

   \institute{
    Max-Planck-Institut für Astronomie, Königstuhl 17, 69117 Heidelberg, Germany
    \and Fakultät für Physik und Astronomie, Universität Heidelberg, Im Neuenheimer Feld 226, 69120 Heidelberg, Germany \and
    Department of Astronomy, California Institute of Technology, Pasadena, CA 91125, USA \and
    Homer L. Dodge Department of Physics and Astronomy, University of Oklahoma, 440 W. Brooks Street, Norman, OK 73019, USA \and 
    JILA, University of Colorado and National Institute of Standards and Technology, 440 UCB, Boulder, CO 80309-0440, USA
}
    \date{Submitted 7 October 2025 / Accepted 26 February 2026}
    
  \abstract
    {We present a rigorous identification of candidates for dormant black holes (BHs) and neutron stars (NSs) in binaries using summary statistics from \textit{Gaia} Data Release 3 (DR3), rather than full orbital solutions. Although \textit{Gaia} astrometric orbits have already revealed a small sample of compact object binaries, many systems remain undetected due to stringent quality cuts imposed on the published orbits. 
    Using a forward-modelling framework that simulates \textit{Gaia} observables, in particular the re-normalised unit weight error (\texttt{ruwe}) and radial velocity (RV) scatter, we infer posterior distributions for companion mass and orbital period via Markov chain Monte Carlo (MCMC) sampling, marginalising over nuisance orbital parameters. We validate our approach by comparing the predicted masses and periods against full orbit solutions from DR3, and by successfully recovering known compact object binaries as promising candidates.
    The method is best suited for systems with red giant primaries, which have more reliable \textit{Gaia} RV scatter and a light centroid more likely dominated by one component, compared to main-sequence stars, and they are less likely to be triples with short-period inner binaries, which produce confounding signatures. 
    We applied the method to three million giants and identify 389 systems with best-fit companion masses $\gtrsim 3\,\mathrm{M}_\odot$. Recovery simulations suggest our selection method is substantially more sensitive than the DR3 non-single-star catalogue, particularly for binaries with periods below 1 year and above $\sim 6$ years. 
    These candidates represent promising targets for spectroscopic follow-up and \textit{Gaia} DR4 analysis to confirm the presence of compact objects and build a more complete census of the Galactic BH and NS population. Candidate main-sequence stars with massive companions face a larger set of confounding effects. Therefore, we present an analogous catalogue of 279 additional main-sequence candidates only as an appendix.
    }%


   \maketitle
%

\section{Introduction}
Stellar-mass black holes (BHs) are compact remnants of massive stars, formed through the gravitational collapse of their cores at the end of stellar evolution. Their masses, orbits, and binary companions encode crucial information about  their formation pathways, from the final stages of massive star evolution to supernova mechanisms and natal kicks. Their subsequent evolution may make them X-ray binaries or gravitational wave sources. Understanding the population demographics of stellar-mass BHs is thus essential for benchmarking theoretical models. However, our observational picture remains largely incomplete, with poorly constrained empirical connections between stellar progenitors (in terms of mass, metallicity, and binarity) and the resulting BH properties.

Until recently, BHs had  only been detected in two subpopulations: X-ray binaries and gravitational wave events. Accreting BHs in close binaries can be identified through their high-energy X-ray emission, and there have been approximately two dozen dynamically confirmed Galactic systems to date \citep[e.g.][]{Remillard_McClintock2006,Casares_Jonker2014,Corral-Santana+2016}. Merging BHs in double compact systems are detected via gravitational wave emission;  more than 200 candidate events are reported in the latest LIGO/VIRGO/KAGRA observing run \citep{LIGO2023}.
However, both of these detection methods trace rare evolutionary pathways, and therefore sample only a tiny fraction of the total BH population \citep[$\ll1\%$; e.g.][]{Wiktorowicz+2019}. Gravitational wave sources are observed only at cosmological distances and form through specific binary evolution scenarios, while only a small subset of BHs in binaries are actively accreting and are luminous in X-rays. As a result, both methods provide a highly biased and incomplete view of the overall BH population.

A much larger fraction of BHs is expected to exist in wide binaries with luminous companions or as isolated objects. While gravitational microlensing provides a potential pathway for identifying isolated BHs, it requires rare alignments and has yielded very few good candidates to date \citep[e.g.][]{Lam+2022, Sahu+2022,Kruszynska+2024,Howil+2025}. 
Non-accreting dormant BHs in binaries with stellar companions thus represent a particularly promising alternative avenue to expand the census of Galactic BHs. These systems are expected to be substantially more numerous than their accreting or merging counterparts, and can be detected through astrometric, kinematic, or photometric signatures induced by the BH on the companion \citep[e.g.][]{Shahaf+2023,Jayasinghe+2023,Gomel+2023}. The presence of a luminous companion further offers a unique opportunity to probe the properties and formation history of the BH through the observable characteristics of the surviving star.

The third data release (DR3; \citealt{GaiaDR3}) of the \textit{Gaia} mission \citep{GaiaMission2016} has opened up new possibilities to identify dormant BHs in binaries \citep[e.g.][for a review on the \textit{Gaia} binary renaissance]{El-Badry2024d}. DR3 includes an extensive non-single-star (NSS) catalogue of binary orbits, derived from astrometric and spectroscopic measurements, with 169,227 and 220,372 published orbit solutions, respectively \citep{Halbwachs+2023,Gosset+2025}. This dataset has already enabled the discovery of the first Galactic dormant BHs \citep{El-Badry+2023b, Chakrabarti+2023, El-Badry+2023, Panuzzo+2024, Wang+2024}, as well as a dozen candidate neutron star (NS) companions \citep{El-Badry+2024b, El-Badry+2024a}, all vetted by spectroscopic follow-up. These systems feature compact objects in wide orbits (typically $P \gtrsim 1$\,yr) around low-mass stars. Their discovery underscores the power of \textit{Gaia} to reveal a previously hidden population of compact remnants in the solar neighbourhood.

\begin{figure}
\includegraphics[width=\columnwidth,angle=0]{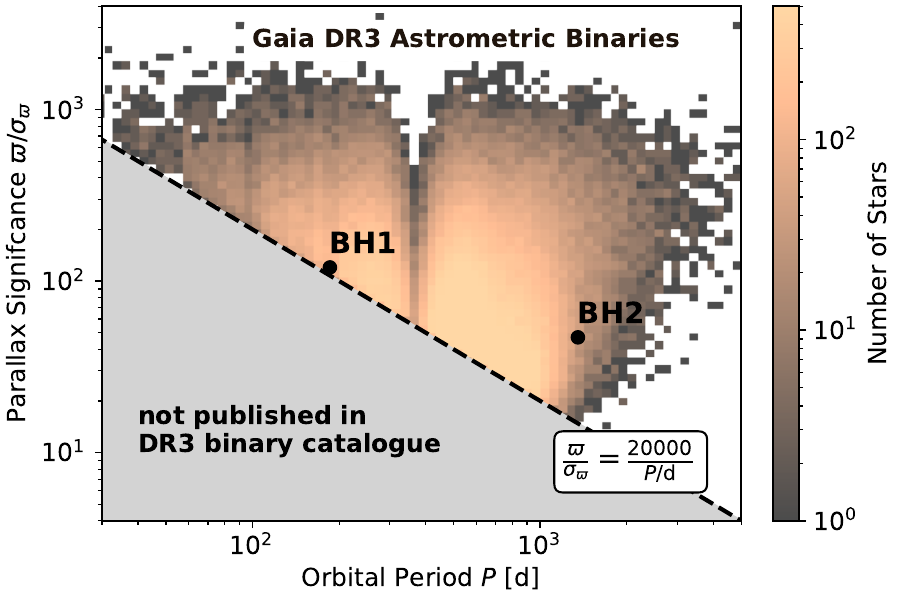}
\caption{Cut on period vs parallax over parallax uncertainty for binaries in the DR3 astrometric catalogue and the dormant BH binaries \textit{Gaia} BH 1, 2 \citep{Chakrabarti+2023,El-Badry+2023b,El-Badry+2023}. The strict quality cuts for \textit{Gaia} DR3 excised many binaries (some with BH companions) from the published \textit{Gaia} NSS catalogue.}
\label{fig:nss_quality_cuts} 
\end{figure}

However, the NSS catalogues with orbit solutions in \textit{Gaia} DR3 were subject to stringent quality cuts \citep[e.g.][]{Halbwachs+2023}. For example, only astrometric binary solutions exceeding a strict signal-to-noise (S/N) threshold in parallax were included (Fig.~\ref{fig:nss_quality_cuts}). Notably, even the recently identified \textit{Gaia} BH3 \citep{Panuzzo+2024},  a $33\,\mathrm{M}_\odot$ BH orbiting a giant star, did not meet the S/N criteria for inclusion in the astrometric binary catalogue. Similarly, the candidate mass-gap BH \textit{Gaia} DR3 3425577610762832384 \citep{Wang+2024} did not receive an astrometric solution. These cases highlight that many more BH and NS binaries may exist within close proximity to the Sun, but remain unidentified due to the quality cuts for orbit solutions applied in DR3.

In addition to orbital solutions, \textit{Gaia} DR3 provides time-averaged parameters and variability metrics for a much larger sample of stars. 
These statistics include the radial velocity (RV) uncertainty $\sigma_\mathrm{RV}$ \citep{Katz+2023}, the re-normalised unit weight error \texttt{ruwe} from the astrometric solution \citep{Lindegren2018}, and the photometric variability in the $G$, $G_\mathrm{BP}$, and $G_\mathrm{RP}$ bands $\sigma_\mathrm{phot}$ \citep{Riello+2021}. The light centroid motion (in position or in RV) of unresolved binaries can produce excess scatter in these observables, providing indirect signatures of binarity even in the absence of a published orbit. These metrics are typically derived from 20–80 observational epochs and are available for 33 million stars \citep[][]{Katz+2023}.

The light centroid's orbital motion affects the stellar position on the sky, degrading the fit of the single-star astrometric model and leading to elevated \texttt{ruwe} values and astrometric excess noise. Spectroscopic binaries may exhibit inflated RV uncertainties and larger RV amplitudes, while close systems may show enhanced photometric variability due to ellipsoidal modulation or eclipses. These correlations have been widely exploited to identify unresolved binaries. In particular, \texttt{ruwe} has been shown to trace binarity across various regions of the colour-magnitude diagram \citep[CMD;][]{Belokurov+2020, Penoyre+2022b, Korol+2022}, and analytical models have been developed to predict its behaviour in binaries \citep[e.g.][]{Penoyre+2022a, Andrew+2022, Castro-Ginard+2024}. For systems with orbital periods shorter than the \textit{Gaia} mission baseline, the expected \texttt{ruwe} signal approximately scales with the angular size of the photocentre orbit \citep{Stassun_Torres2021}, making \texttt{ruwe} most sensitive to nearby systems with intermediate periods ($P \sim 10^3$ days) and massive faint companions.

While each summary statistic on its own provides only weak constraints and may be influenced by other astrophysical variability (e.g. pulsations, spots, or multiple luminous stars), their joint modelling increases diagnostic power. Previous work by \citet{Andrew+2022} combined astrometric, photometric, and spectroscopic variability indicators to search for compact object companions. Using analytic estimates of excess noise, they identified promising candidates in DR3 data. A key challenge in their analysis was contamination from hierarchical triples, where large RV amplitudes and astrometric motion can arise from distinct companions in inner and outer orbits.

Building on these efforts, we leveraged the \textit{Gaia} DR3 dataset and adopted a forward model of the entire \textit{Gaia} observing pipeline \citep{El-Badry+2024c}. This approach allowed us to interpret the observed combination of astrometric, spectroscopic, and photometric variability in a consistent and quantitative framework. Our model fits these observables simultaneously to constrain the mass $M_\mathrm{CO}$ and orbital period $P$ of possible binary companions. By selecting systems with high inferred companion masses and mass ratios, we  recover known compact object binaries and identify new candidates that may host dormant BHs or NSs. By focussing on evolved stars, where present or past large radii tend to exclude tight inner binaries, we reduced the contamination from unresolved higher-order multiples.

Section~\ref{sec:forward_model} gives an overview of the forward model framework we use to model \textit{Gaia} summary diagnostics. In Sect.~\ref{sec:parameter_inference} we describe our approach to constraining binary parameters using these observables. The application of this method to evolved stars in \textit{Gaia} DR3 is presented in Sect.~\ref{sec:rgb_sample}, along with the resulting catalogue of candidate binaries with compact companions. We discuss potential sources of contamination in Sect.~\ref{sec:discussion}, and summarise our findings in Sect.~\ref{sec:conclusion}.

\section{Forward modelling \textit{Gaia} summary statistics}
\label{sec:forward_model}

The standard single-star model fit to \textit{Gaia} sources outside the NSS catalogues assumes that the apparent movement of a star on the sky can be fully described by parallax and proper motion, with a constant RV over time. These assumptions break down in binary systems, where orbital motion introduces additional variability. This variability manifests itself as excess noise, leading to poorer fits under the single star model and inflated uncertainties in the astrometric and spectroscopic parameters.

In this section we describe our forward-modelling framework, which predicts the expected summary statistics reported by \textit{Gaia} as a function of binary parameters. This modelling step is then used to solve the inverse problem: given a set of observed summary statistics, what physical parameters and orbits can explain them?

\begin{figure}[ht!]
    \centering
    \begin{minipage}[t]{\columnwidth}
        \centering
        \includegraphics[width=\columnwidth]{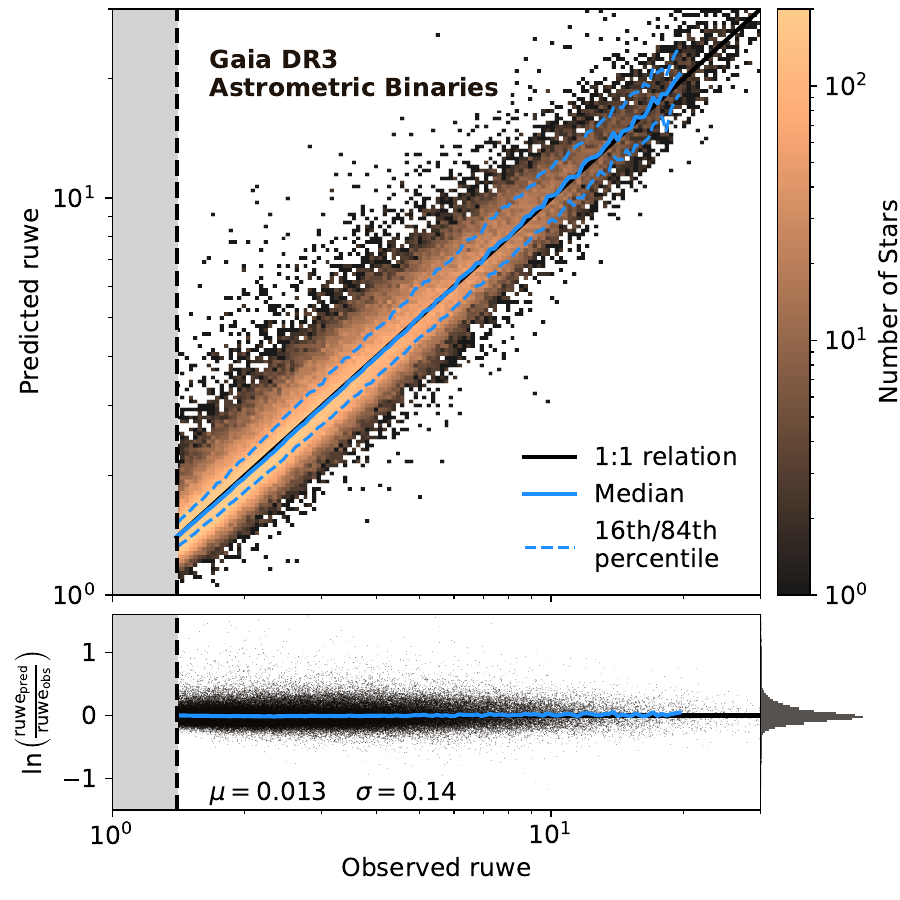}
    \end{minipage}
    \begin{minipage}[t]{\columnwidth}
        \centering
        \includegraphics[width=\columnwidth]{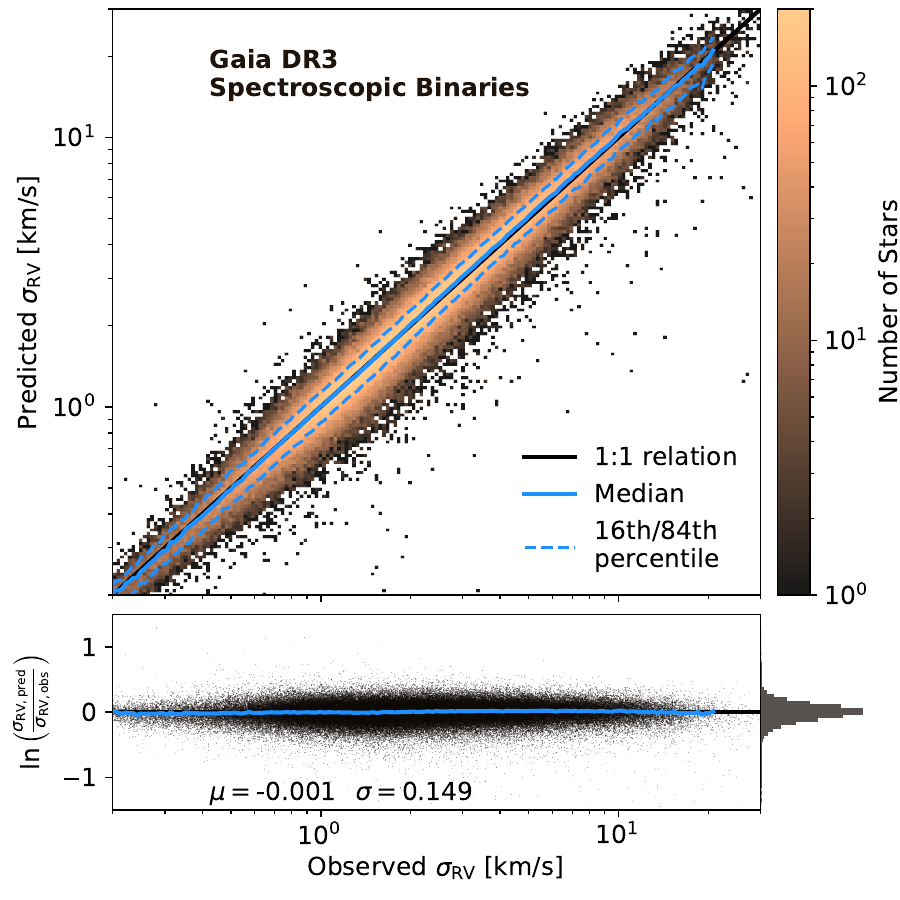}
    \end{minipage}
    \caption{Predicted vs catalogued \textit{Gaia} DR3 summary statistics for known binaries. The predictions are based on forward modelling of their orbital parameters. Top panel: Comparison between predicted and catalogue values of the astrometric goodness of fit of the single-star astrometric solution, \texttt{ruwe}, for binaries with published orbital solutions. Bottom panel: Predicted vs observed RV errors for spectroscopic binaries (SB1) in \textit{Gaia} DR3. In both plots the blue line shows the median trend, while the dashed blue lines indicate the 16th and 84th percentiles. The 1:1 relation is marked in black. The lower subpanels show the logarithmic residuals between observed and predicted values; the side histograms illustrate their normalised distribution. The mean bias ($\mu$) and scatter ($\sigma$) of the residuals are indicated.} 
    \label{fig:forward_model}
\end{figure}

\subsection{Astrometry}
\label{sec:methods_astrometry}

We used the generative forward model \texttt{gaiamock} \citep{El-Badry+2024c} to simulate \textit{Gaia} astrometric observations for unresolved binaries and predict the resulting summary statistics, particularly the re-normalised unit weight error \texttt{ruwe}. \texttt{gaiamock} generates realistic epoch astrometry based on the \textit{Gaia} scanning law, and the source parameters, including position ($\alpha, \delta$), parallax $\varpi$,  proper motions ($\mu_\alpha, \mu_\delta$), $G$ band magnitude, flux ratio, and component masses ($M_*, M_\mathrm{CO}$). Alternatively,  \texttt{gaiamock}  can take as input the semi-major axis of the photocenter $a_0$ and the binary orbital elements: period $P$, time of periastron passage $T_\mathrm{peri}$, eccentricity $e$, inclination $i$, argument of periastron $\omega$, and longitude of ascending node $\Omega$. The model uses an empirical prescription for the epoch astrometry errors, calibrated on well-behaved single stars.

From the simulated epoch astrometry, \texttt{gaiamock} fits a five-parameter single-star model, predicts the resulting along-scan displacements, and computes the corresponding chi-squared ($\chi^2$) statistic of the residuals. While a minority of \textit{Gaia} DR3 sources are fitted with a six-parameter astrometric model \citep{Lindegren+2021}, these constitute a small fraction of our sample and were treated identically in the forward modelling with \texttt{gaiamock}. Following \textit{Gaia} conventions, the per-epoch uncertainties are inflated based on the goodness of fit, and \texttt{ruwe} is approximately the square root of this reduced chi-squared \citep{El-Badry2025}:

\begin{equation}
    \texttt{ruwe} = \sqrt{\frac{\chi^2}{\nu (1-\frac{2}{9\nu})^3}} \approx \sqrt{\chi^2/\nu}\,.
\end{equation}
Here $\nu = \mathtt{astrometric\_n\_good\_obs\_al} - m$ is the number of degrees of freedom, defined as the difference between the number of observations in the along-scan direction and the number of free parameters; $m=5$ for the five-parameter single-star model.
To validate our astrometric model, we compared predicted \texttt{ruwe} values with those reported in \textit{Gaia} DR3 for sources with published astrometric orbits \citep{Halbwachs+2023, GaiaDR3}. These systems are well suited for validation, as their orbital parameters are fully constrained, enabling accurate forward modelling of their astrometric motion (this is in contrast to systems with only spectroscopic orbits, where, for example, orbital orientation is unconstrained, which would dominate the scatter in predicted \texttt{ruwe}).

We used binaries from the \texttt{two\_body\_orbit} catalogue with either astrometric-only (\texttt{Orbital}) or combined astrometric and spectroscopic (\texttt{AstroSpectroSB1}) solutions and selected 158,690 high-quality orbits with $\texttt{goodness\_of\_fit} < 10$ for $G \leq 13$ and $< 5$ for $G > 13$ \citep[see][]{El-Badry+2023b}. Figure~\ref{fig:forward_model} shows that the predicted and catalogued \texttt{ruwe} values exhibit a tight correlation with negligible systematic offset, consistent with previous results \citep{El-Badry2025}. A minute fraction of outliers show predicted \texttt{ruwe} values larger than observed; these do not exhibit a clear correlation with any single orbital parameter. The scatter in the logarithmic ratio $\ln \left(\texttt{ruwe}_\text{pred} / \texttt{ruwe}_\text{obs}\right)$ is approximately constant $\epsilon_{\texttt{ruwe}} = 0.14$ throughout the \texttt{ruwe} range. We adopt this value as the fractional uncertainty in our \texttt{ruwe} predictions.

\subsection{Spectroscopy}
\label{sec:methods_spectroscopy}

As a spectroscopic variability indicator, we forward model the formal uncertainty on the median RV reported in the \textit{Gaia} DR3 catalogue.
For a star in a Keplerian binary, the RV varies over time according to
\begin{equation}
    v_{\mathrm{rad}}(t) = v_{\mathrm{z}} + K \left(\cos(\omega + \nu(t)) + e\cos(\omega)\right)\,, \label{eq:rv_orbit_model}
\end{equation}
where $\nu$ is the true anomaly, $e$ is the eccentricity, and $v_z$ is a constant velocity offset. The RV semi-amplitude $K_*$ of the primary star induced by the companion is a function of the companion mass, orbital period, eccentricity, and inclination:
\begin{align}
        K_* &=\frac{M_\mathrm{CO}}{(M_*+M_\mathrm{CO})^{\frac{2}{3}}} \, \left(\frac{2\pi G} {P}\right)^{\frac{1}{3}}\,{(1-e^2)^{-\frac{1}{2}}}\,\sin i\,.  \label{eq:rv_amplitude}
    \end{align}
The RV amplitude increases with higher companion masses and smaller separations of the stars.

We use \texttt{gaiamock} to compute epoch-level RVs by sampling Eq.~\ref{eq:rv_orbit_model} at realistic \textit{Gaia}-like observation times. Because the precise spectroscopic scanning law is not public, we approximate the RV observing epochs using a random subset of \texttt{rv\_nb\_transits} times from the observing dates given by the astrometric scanning law. While not exact, this approximation appears sufficient for our purposes: with a median of $\sim$18 transits per star in DR3 \citep{Katz+2023}, binaries with periods shorter than or comparable to the \textit{Gaia} baseline are typically sampled well enough that the specific observing times have little effect on the variability statistics.

To model the reported RV uncertainty, we relate it to the standard deviation of the epoch RVs, $\tilde\sigma_\mathrm{RV}$, using the expression
\begin{multline}
\texttt{radial\_velocity\_error} = \\
\sqrt{\left(\sqrt{\frac{\pi}{2N}} \, \tilde\sigma_\mathrm{RV} \right)^2 + \left(0.113\,\mathrm{km\,s}^{-1}\right)^2},
\end{multline}
where $N$ is the number of transits (\texttt{rv\_nb\_transits}), and the additive term accounts for the spectroscopic noise floor from wavelength calibration uncertainties \citep{Katz+2023}. We compute the standard deviation from the simulated RVs and apply this transformation to predict the expected \texttt{radial\_velocity\_error}, which we refer to as RV uncertainty $\sigma_\mathrm{RV}$.

This model applies only to bright stars with $\texttt{grvs\_mag} \lesssim 12$ and $\texttt{rv\_method\_used} = 1$, for which individual epoch RVs are extracted and analysed. For fainter stars, \textit{Gaia} uses a combined cross-correlation function rather than individual RV epochs, which limits the interpretability of variability statistics. Our approach is best suited to cool stars and giants, for which \textit{Gaia} obtains precise epoch RVs. It is probably less effective for hot or rapidly rotating stars, whose RVS spectra are dominated by shallow, broadened Paschen lines, resulting in significantly lower RV precision \citep{Katz+2023}. 

To validate our spectroscopic variability model and empirically calibrate its uncertainties, we tested it on known spectroscopic binaries from the \textit{Gaia} NSS catalogue, selecting systems with \texttt{nss\_solution\_type} = \texttt{SB1}. Because our predictions included only the RV variability induced by binary motion, neglecting measurement errors and additional noise sources, they would underestimate the reported RV uncertainties. To account for this, we applied a uniform correction factor derived from the observed biases in the validation sample and scaled the predicted RV uncertainties by a factor of 1.07. Figure~\ref{fig:forward_model} compares our predicted vs the catalogued RV uncertainties. The tight linear correlation confirms that the model captures the key trends in this metric.
As in the astrometric case, the scatter in the logarithmic residuals is approximately constant. We therefore adopt fractional uncertainties of $\epsilon_\mathrm{RV} = 0.15$.

\subsection{Photometry}
\label{sec:methods_photometry}

To model photometric variability, we adopt an analytical prescription for the amplitude of ellipsoidal variations, that is, the brightness changes arising from tidal distortion of a star in a close binary system \citep{Morris_Naftilan1993}. These effects are strongest at short orbital periods and diminish rapidly when the binary separation is much larger than the radii of the component stars. Our aim is not to capture the variability in detail but to estimate its amplitude to order-of-magnitude precision, sufficient to exclude short-period binaries from our candidate sample.

\citet{Morris_Naftilan1993} derive an analytical model for ellipsoidal modulation in circular orbits, expressed as a power series in $R_*/a$, the ratio of the primary star’s radius to the system’s semimajor axis. We use the dominant second-harmonic term from their Eq. (1), incorporating the correction factor from \citet{Gomel+2021}. Under the assumption that the primary star dominates the flux, the semi-amplitude of the photometric variability $A_\mathrm{phot}$ can then be expressed as a function of the binary mass ratio $q$, inclination angle, and Roche-lobe filling factor of the primary (which depends on $R_*$, $M_*$, $q$, and $a$):
\begin{align}
        A_{\mathrm{phot, pred}}  &= \,c_2 \left(\alpha_{2} \left(\frac{R_*}{a}\right)^3  q \sin^2 i +\, \beta_{2} \left(\frac{R_*}{a}\right)^5 q \left(6 \sin^2 i - 7 \sin^4 i\right) \right)
    \end{align}

The coefficients $\alpha_{2}$ and $\beta_{2}$, along with the correction factor $c_2$ \citep[see Eqs. 2 and 11 in][]{Gomel+2021}, account for limb and gravity darkening, and depend on the stellar effective temperature and surface gravity. We adopted approximate coefficient values consistent with a star with properties of the companion to \textit{Gaia} BH~2 \citep{Claret+2011, El-Badry+2023b}, which is adequate given our focus on detached binaries containing cool stars with convective envelopes.

As such, this simplified model functions primarily as a filter: By flagging systems with predicted $A_\mathrm{phot}$ above a threshold amplitude, it allows us to conservatively exclude close binaries and hierarchical triples from our candidate list (Sect.~\ref{sec:cleaning}). The predicted variability amplitude is reflected in the observed scatter in the \textit{Gaia} $G$ band photometry, parametrised by
\begin{equation}
A_{\mathrm{phot, obs}} = \frac{ \sqrt{2} \sqrt{\texttt{phot\_g\_n\_obs}}}{\texttt{phot\_g\_mean\_flux\_over\_error}} \,.
\end{equation}

\section{Binary parameter inference}
\label{sec:parameter_inference}

\begin{figure*}[t]
\centering
\includegraphics[width=\textwidth]{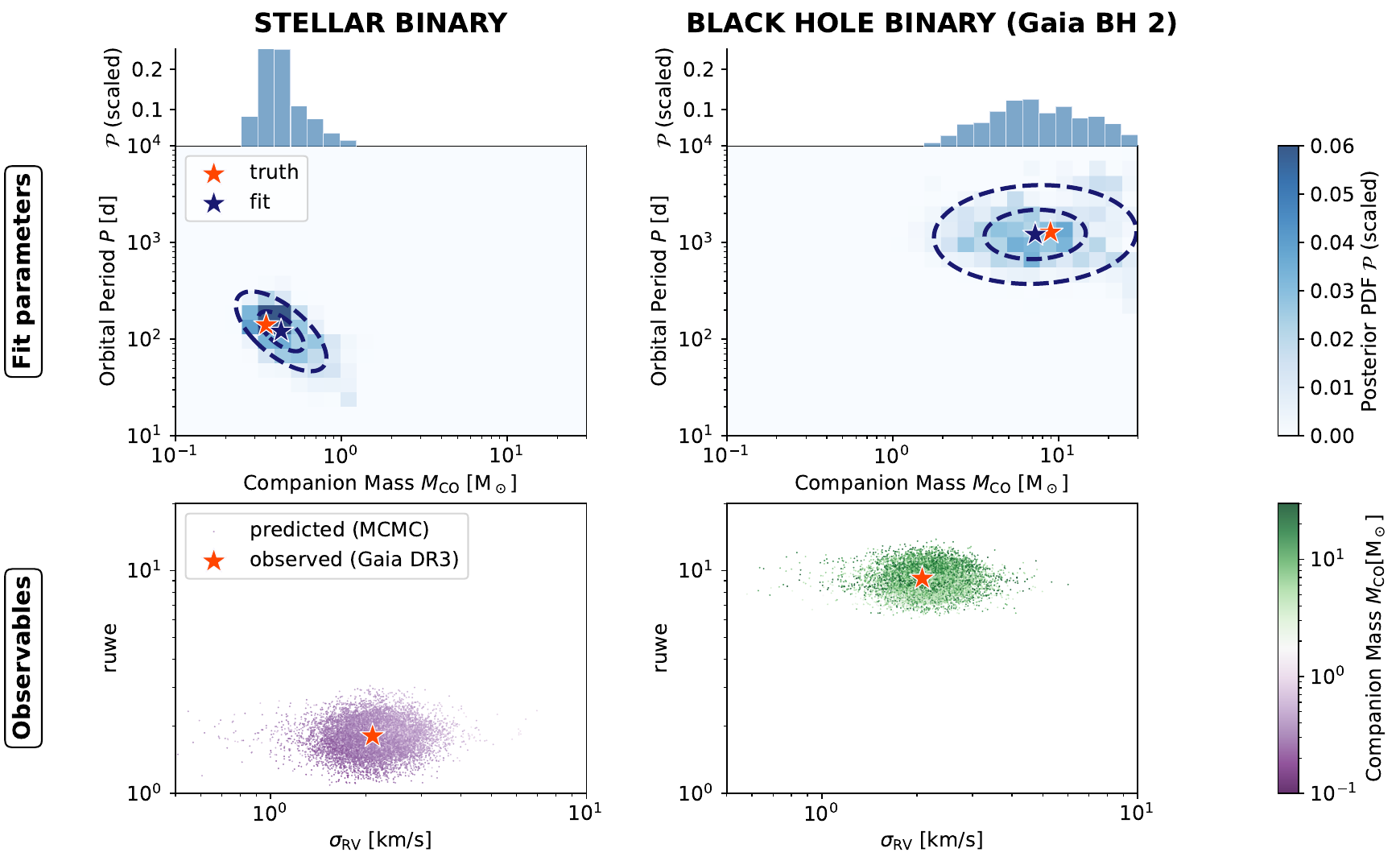}
\caption{Posterior inference for two example binary systems using our forward-modelling approach: a luminous stellar binary (DR3~4609711364265062016, left panels) and a compact-object binary, \textit{Gaia} BH~2 (DR3~5870569352746779008, right panels). Top row: Posterior samples in the space of companion mass vs orbital period, along with the marginalised posterior distribution of the companion mass. The orange star symbols mark the true (injected) parameters of each system; the blue stars denote the inferred best-fit parameters (posterior-weighted mean). The dashed ellipses indicate the 1$\sigma$ and 2$\sigma$ confidence contours derived from the covariance matrix. Bottom row: Predicted observables (\texttt{ruwe} vs RV uncertainty $\sigma_\text{RV}$) for the same posterior samples, colour-coded by the companion mass. The observed values are again shown as orange stars. In the stellar case, the low-mass companions reproduce the observed signatures well; in the BH case, a high companion mass is required to explain the elevated \texttt{ruwe} and moderate $\sigma_\text{RV}$, consistent with the true binary configurations. Much of the variance in predicting the observables results from the random sampling of RV epochs, which can produce varying observables also for the same binary parameters.}
\label{fig:comparison_fit}
\end{figure*}

In Sect.~\ref{sec:forward_model} we showed that for known binary parameters, we can reliably predict \textit{Gaia} observables, in particular the re-normalised unit weight error (\texttt{ruwe}) and the RV uncertainty ($\sigma_\mathrm{RV}$). We now address the inverse problem: How well can one constrain binary parameters from these summary diagnostics using a forward-modelling Bayesian framework. 

We adopted a Markov chain Monte Carlo (MCMC) approach, using the \texttt{emcee} sampler \citep{Foreman-Mackey+2013}, to sample the posterior distribution of binary parameters given the observed \textit{Gaia} summary statistics. Specifically, we inferred the posterior probability distribution of the parameter vector
\[
\vec{\theta} = \left(P, M_\text{CO}, \cos i, e, \Omega, \omega, T_\text{p}, \varpi\right),
\]
given the observables
\[
\vec{d} = \left(\texttt{ruwe}, \sigma_\text{RV}\right),
\]
and assuming known stellar parameters,
\[
\vec{s_*} = \left(M_*, T_{\text{eff,}*}, R_*, G, \alpha, \delta\right).
\]

The orbital parameters cannot be uniquely constrained by the available observables alone. However, we assessed whether meaningful constraints can still be obtained on the most relevant parameters by marginalising over the other ones. Our primary parameters of interest are the orbital period ($P$) and the unseen companion mass ($M_\text{CO}$). The remaining orbital parameters -- inclination ($i$), eccentricity ($e$), longitude of the ascending node ($\Omega$), argument of periastron ($\omega$), time of periastron passage ($T_\text{p}$), and parallax ($\varpi$) -- were treated as nuisance parameters required for forward modelling but marginalised in the final inference. Although parallaxes are provided in the \textit{Gaia} DR3 catalogue, we treated $\varpi$ as a free parameter. This allowed parallax uncertainties to be consistently propagated into the inferred orbital parameters and companion masses, which is particularly important for systems with low parallax significance.

The log-likelihood function is defined as
\begin{multline}
\log \mathcal{L}(\vec{\theta}) =
-\frac{1}{2} \ln(2\pi\epsilon^2_{\texttt{ruwe}})
-\frac{1}{2} 
\left(\frac{ \ln \texttt{ruwe}_\mathrm{obs} - \ln \texttt{ruwe}_\mathrm{pred}(\vec{\theta}) }{\epsilon_{\texttt{ruwe}}} \right)^2 \\ 
-\frac{1}{2}\ln(2\pi\epsilon^2_{\sigma_\text{RV}})
-\frac{1}{2} \left(\frac{ \ln \sigma_{\text{RV,obs}} - \ln \sigma_{\text{RV,pred}}(\vec{\theta}) }{\epsilon_{\sigma_\text{RV}}}\right)^2
,
\end{multline}
where `obs' denotes the catalogued summary statistics and `pred' the model predictions for a given parameter set $\vec{\theta}$. 
The likelihood measures discrepancies in units of scatter, given by $\epsilon_{\texttt{ruwe}}$ and $\epsilon_{\sigma_\text{RV}}$, and penalises them in log-space. The use of logarithmic quantities is empirically motivated; forward modelling of NSS systems shows that the combined scatter -- from intrinsic variation and model uncertainty -- is constant in log-space and can be approximated by a Gaussian distribution (see Fig.~\ref{fig:forward_model}). Expressing the likelihood in terms of $\ln \mathtt{ruwe}$ and $\ln \sigma_\mathrm{RV}$ also ensures a symmetric treatment of relative over- and under-predictions.

We adopted log-uniform priors for period and companion mass, and uniform priors for all other orbital parameters over physically motivated ranges: $\log P/\mathrm{d} \in [1, 4]$, $\log M_\text{CO}/M_\odot \in [\log(0.05), \log (30)]$, $\cos i \in (0, 1]$, $e \in [0, 0.95]$, $\Omega, \omega \in [0, 2\pi)$, and $T_p/\text{d} \in [0,P]$. For the parallax, we adopted a normal prior centred on the \textit{Gaia} DR3 value after applying the global zero-point correction of \citet{Lindegren+2021b}. The prior standard deviation was set by the reported parallax uncertainty, inflated as a function of \texttt{ruwe} following the prescription of \citet{El-Badry2025}, to account for underestimated astrometric errors in unresolved binaries. The flux ratio was fixed to zero, which is appropriate for dark companions. For luminous companions, a non-zero flux ratio would reduce the observed \texttt{ruwe}, leading to underestimated periods and masses. This bias is acceptable since we target dark, massive companions, and for luminous giants, the flux contribution from stellar companions is typically negligible.

We sampled posterior distributions using 32 walkers and 2,500 steps with a 500 step burn-in phase. Convergence assessment is challenging due to the underconstrained problem and high shot noise (e.g. from stochastic \textit{Gaia} RV observation date sampling). We empirically evaluated convergence on a subset of stars by verifying that: (a) randomly initialised walkers converge to consistent posterior distributions, (b) marginal distributions of $P$ and $M_\text{CO}$ remain stable between the second and final thirds of the chain, and (c) increasing chain length by a factor of 10 does not significantly alter results. The inference is computationally efficient, requiring approximately 10 seconds per star on a single CPU core. We report posterior-weighted mean values of $P$ and $M_\text{CO}$, marginalised over nuisance parameters.

Figure~\ref{fig:comparison_fit} illustrates the results of this MCMC fitting procedure for two test binary systems: a stellar binary (left panels) and a system containing a compact object (right panels), specifically \textit{Gaia} BH~2 \citep{El-Badry+2023b}. The upper panels display the marginalised posterior probability distribution function (PDF) in the plane of orbital period vs companion mass. The orange star denotes the true input parameters of each system, while the blue star marks the inferred best-fit parameters, defined as the posterior-weighted mean. The dashed ellipses show 1-$\sigma$ and 2-$\sigma$ confidence contours, derived from the posterior covariance matrix. Above each panel, we show the marginalised 1-D posterior PDF of the companion mass, normalised so that the sum of all posterior weights is unity. The bottom panels show the predicted observables \texttt{ruwe} and $\sigma_\text{RV}$ for each MCMC sample, colour coded by the corresponding companion mass. The orange star shows the observed values (from the \textit{Gaia} source catalogue).

The first system (\textit{Gaia} DR3 4609711364265062016) appears in the NSS catalogue as a \texttt{AstroSpectroSB1} binary with an orbital period of 141~d. The \textit{Gaia} binary mass catalogue assigns it a primary mass of $\sim 1.1\,M_\odot$ and a companion mass of $\sim 0.35\,M_\odot$. The observed values of \texttt{ruwe} and RV uncertainty are modest ($\texttt{ruwe} = 1.81$, $\sigma_\text{RV} = 2.10$\,km\,s$^{-1}$), consistent with a luminous stellar companion. Our inference correctly recovers a low-mass companion in a short-period orbit ($P_\text{fit} = 120^{+80}_{-50}\,\text{d},\,\, M_\text{CO, fit} = 0.4\pm 0.2\,\text{M}_\odot$). Small mismatches between the true and inferred values and large uncertainties are expected, given the degeneracies in orbital orientation and eccentricity.

The second example is \textit{Gaia} BH~2 (\textit{Gaia} DR3 5870569352746779008), a giant star orbited by a $\sim 9\,\text{M}_\odot$ BH in a long-period  eccentric orbit \citep[$P \approx 1277$\,d, $e \approx 0.53$;][]{El-Badry+2023b}. It has a high \texttt{ruwe} value of $9.22$, with moderate RV uncertainty ($\sigma_\text{RV} = 2.08$\,km\,s$^{-1}$). Matching both observables requires a high companion mass, and our inference yields a posterior peak at $P_\text{fit} = 1200^{+1000}_{-600}\,\text{d}$ and $M_\text{CO, fit} = 7^{+8}_{-4}\,\text{M}_\odot$, consistent with the true values and a dark remnant companion. For systems with orbital periods longer than the DR3 baseline, there is a degeneracy between period and companion mass; a longer-period, more massive companion can produce similar observables (\texttt{ruwe}, $\sigma_\text{RV}$) as a shorter-period, less massive one due to limited orbital coverage. We capped the prior at $30\,\text{M}_\odot$, which constrained the posterior to a physically motivated BH mass regime, but also truncated the distribution. For candidates like BH~2 that lie near this boundary, the inferred companion masses should therefore be interpreted as lower limits under the assumed log-uniform prior.

These examples illustrate that our method is capable of identifying massive compact companions from summary statistics alone and supports its application to the detection of dormant BHs.

\section{A catalogue of RGB+BH candidates}
\label{sec:rgb_sample}

We are now in a position to apply this forward-modelling technique to a sample of evolved stars from \textit{Gaia} DR3. Giant stars offer several advantages in searches for compact companions: (1) with their high luminosities they are likely to outshine their companions, resulting in flux ratios near zero and photocentre motions dominated by the primary; and (2) while stars low on the RGB may still host companions in close ($\sim$ weeks) orbits, the large current or past radii of more evolved giants preclude short-period orbits, reducing contamination from hierarchical triples, a major source of false positives in main-sequence binary searches \citep{Andrew+2022}. A corresponding main-sequence sample is provided in Appendix~\ref{app:ms_candidates}.

\subsection{Sample selection and filtering}

For our initial sample, we queried the \textit{Gaia} DR3 database for sources that meet the following criteria.
\begin{lstlisting}
SELECT *
FROM gaiadr3.gaia_source
WHERE phot_g_mean_mag + 5*log10(parallax/100) < 3 
AND bp_rp > 1
AND parallax > 0
AND ipd_frac_multi_peak = 0
AND rv_method_used = 1
\end{lstlisting}

This yielded a parent sample of 3,375,482 stars. The first two cuts select red and luminous evolved stars in the CMD. Requiring positive parallaxes effectively imposes a distance cut, which is necessary for interpreting and modelling the astrometric data. The fourth cut removes blended sources, and the final requirement ensures that the stars have measured epoch RVs, to which our spectroscopic model applies (Sect.~\ref{sec:methods_spectroscopy}). 

We are interested in stars showing signs of both astrometric and spectroscopic variability. Therefore, we also required
\begin{itemize}
    \item $\mathtt{rv\_amplitude\_robust} > 20$\,km\,s$^{-1}$ and 
    \item $\mathtt{ruwe} > 1.4$\,. 
\end{itemize}
These cuts reduced the sample to 37,421 stars. The RV amplitude threshold roughly corresponds to $\sigma_\mathrm{RV} \gtrsim 2$\,km\,s$^{-1}$.
These thresholds are motivated by two considerations: (a) In broader tests, systems with low RV scatter and low \texttt{ruwe} -- which occupy the lower left region of the bottom panels in Fig.~\ref{fig:comparison_fit} -- were almost exclusively binaries with low-mass companions. Due to the rarity of BHs and the high contamination in this region, we focussed on areas of parameter space where BHs exhibit stronger and more distinct dynamical signatures. (b) Although BHs in wide or low-inclination orbits may be excluded by these cuts, we found that the impact on overall completeness is small for the period range of interest (see Sect.~\ref{sec:completeness}).

To further refine the sample, we cross-matched with the stellar parameter catalogue from \citet{Zhang+2023}, which provides $T_\mathrm{eff}$, $\log g$, and [Fe/H] estimates from BP and RP (XP) spectra for 220 million stars. We retained stars with $T_\mathrm{eff} < 6000$\,K to remove reddened main-sequence stars, and required reliable parameter estimates (\texttt{quality\_flags} < 8). This removed very red supergiants and reduced the sample to 21,028 evolved stars.

\begin{figure}[t]
    \centering
    \includegraphics[width=\columnwidth]{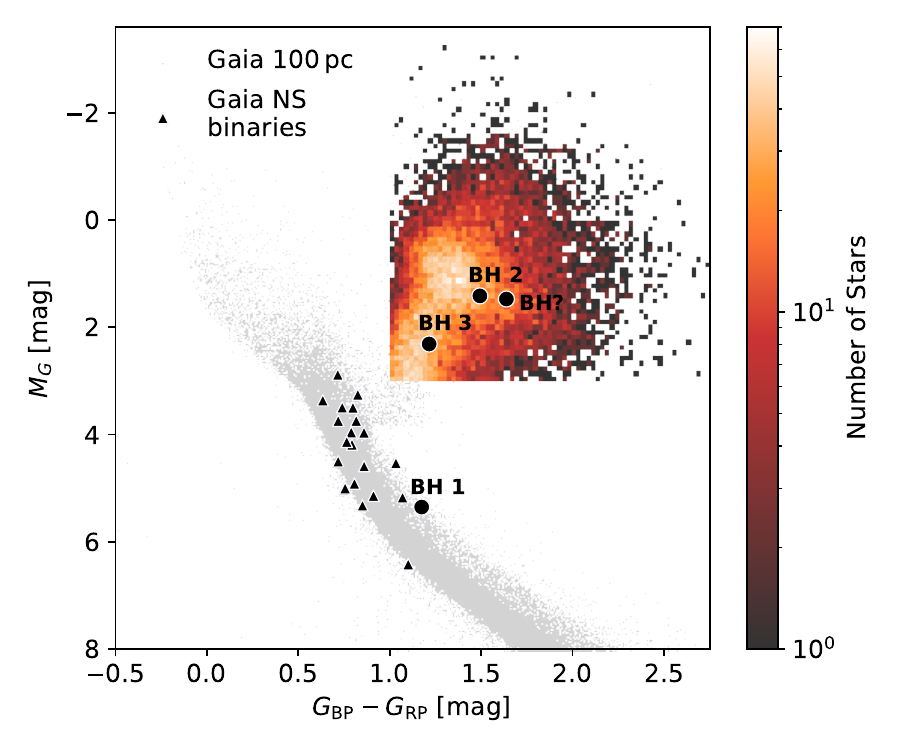}
    \caption{CMD of the RGB candidate sample (2D histogram), overlaid on all stars within 100\,pc from \textit{Gaia} DR3 (grey points). Highlighted are the confirmed BH binaries \textit{Gaia} BH~2 and BH~3, and the candidate LAMOST BH. Neutron star binaries (triangles) and \textit{Gaia} BH~1 are shown for reference, but lie outside our RGB selection.}
    \label{fig:cmd_rgb_sample}
\end{figure}

Figure~\ref{fig:cmd_rgb_sample} shows the CMD of the final sample. We note that the confirmed dormant BHs \textit{Gaia} BH~2 \citep{El-Badry+2023b} and BH~3 \citep{Panuzzo+2024}, along with the candidate mass-gap BH identified from LAMOST \citep{Wang+2024}, are within this selection and are highlighted. \textit{Gaia} BH~1 \citep{El-Badry+2023} and the candidate NS binaries from \citet{El-Badry+2024b} are excluded, as their primary stars are on the main sequence and fainter than the DR3 limit for epoch RVs.

\subsection{Stellar parameter estimation}
\label{sec:stellar_parameters}
Our forward model relies on stellar parameters of the luminous star, particularly its mass, which is essential for interpreting the binary mass function and constraining the mass of the unseen companion. For evolved stars, precise mass measurements are challenging. Asteroseismic measurements or abundance-based estimates (e.g. using photospheric C/N ratios) represent the gold standard but are not available for most stars in our sample.

Instead, we trained a supervised machine learning model to predict stellar masses and radii from stellar parameters and infrared photometry. Specifically, we used the \texttt{XGBoost} gradient boosting algorithm, trained on stars from the APOKASC-3 catalogue \citep{Pinsonneault+2025}, a set of giants with high-precision stellar parameters, asteroseismic masses and radii derived from APOGEE spectra and Kepler light curves. We limited the training sample to stars with mass and radius uncertainties below 0.1\,M$_\odot$ and 0.1\,R$_\odot$, respectively.
The training feature vector included fundamental stellar parameters ($T_\mathrm{eff}$, $\log g$, [Fe/H]) from \citet{Zhang+2023}, along with absolute magnitudes and colours in the near- and mid-infrared from 2MASS (J, H, K$_s$) and WISE (W1, W2) computed using the \textit{Gaia} parallax measurements. Colour indices (J$-$K, W1$-$W2) were also included.
Stellar luminosities were then derived via the Stefan–Boltzmann law from the inferred radii and temperatures. Additional details of the model architecture and validation tests are provided in Appendix~\ref{app:xgboost}.

Figure~\ref{fig:rgb_masses_and_radii} shows the distributions of stellar masses and radii in the final sample. The stellar masses peak slightly above one solar mass, with a median of 1.3\,M$_\odot$, and the median inferred radius is 8.9\,R$_\odot$. The mass distribution is consistent with the expectations for luminous giants in the Galactic disc \citep{Yu+2018,Wu+2019}. We estimated typical mass uncertainties to be $\sim$15\% ($\approx 0.2\,$M$_\odot$), based on cross-validation in the training set.

While not as precise as asteroseismic methods, this approach is scalable and effective for large samples. The stellar radii enter the forward model only to predict photometric variability (from ellipsoidal distortions) and thus do not require high accuracy. The stellar masses enter modelling only to identify compact object companions significantly more massive than the giant itself (i.e. $M_\mathrm{CO} \gg M_*$), rather than to obtain precise estimates for $M_*$. For any promising BH candidates, more detailed spectroscopic and photometric modelling of the primary should be performed in follow-up.

\begin{figure}[t]
    \centering
    \includegraphics[width=\columnwidth]{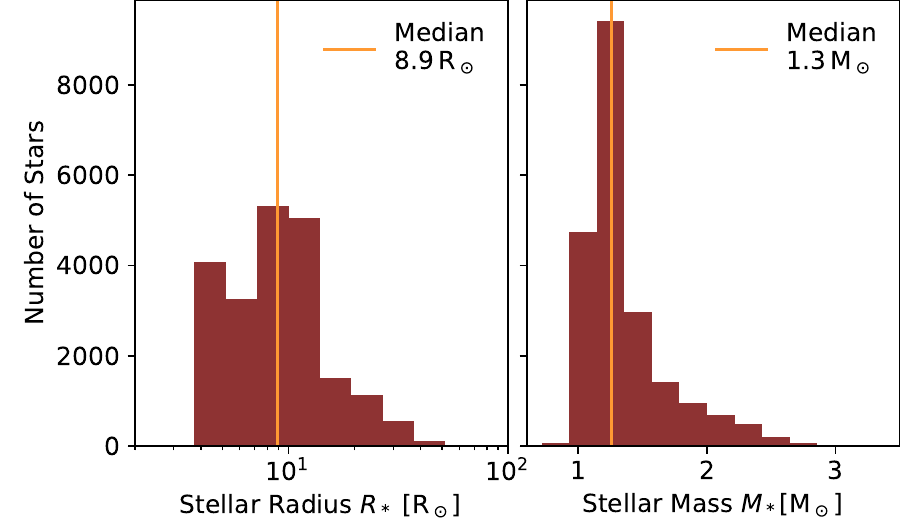}
    \caption{Distributions of inferred stellar masses and radii for the RGB candidate sample. The estimates were obtained using an \texttt{XGBoost} gradient boosting model trained on stellar radii and masses derived from asteroseismology. The giant stars in the sample have median masses of $1.3\,\text{M}_\odot$ and typical radii of $\sim$$10\,\text{R}_\odot$.}
    \label{fig:rgb_masses_and_radii}
\end{figure}

\subsection{Applying the forward model to the RGB sample}

We applied our Bayesian forward-modelling framework to all RGB stars in the filtered candidate sample of 21,028 objects, aiming to constrain their binary parameters. Marginalising over the orbital angles, this yielded best-fit masses and orbital periods for potential compact companions associated with each star.  

\begin{figure}[t]
\includegraphics[width=\columnwidth]{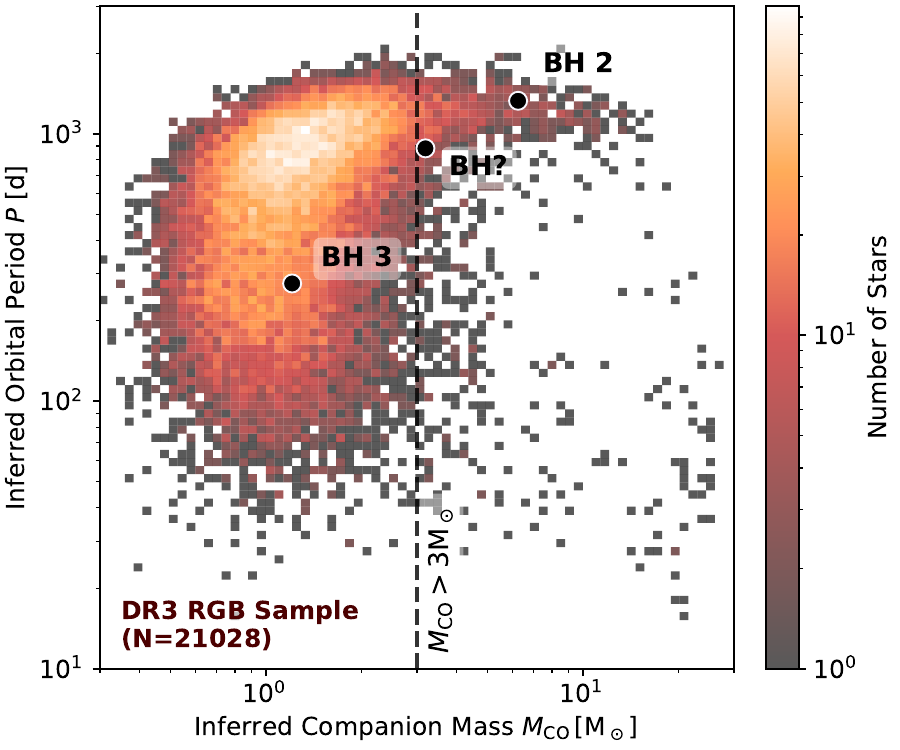}
\caption{Histogram of inferred orbital periods and companion masses for all RGB stars showing excess scatter in RV and position based on DR3 summary diagnostics. The subset of promising candidates for dark companions with high inferred companion masses ($\geq 3\, \mathrm{M}_\odot$) is separated by the dashed vertical line. Also shown are the recovered parameters for \textit{Gaia} BHs 2 and 3, and for the LAMOST BH candidate, all orbiting RGB stars (black markers). Our method successfully recovers \textit{Gaia} BH~2 and the LAMOST binary as promising candidates for RGB + BH systems with high inferred companion masses. \textit{Gaia} BH 3 is not among the identified candidates mostly because our method is less sensitive to objects at very long periods.}
\label{fig:hist_RGB_fit_parameters}
\end{figure}

The distribution of the inferred companion masses and orbital periods for these 21,028 stars is shown in Fig.~\ref{fig:hist_RGB_fit_parameters}. The distribution peaks at low companion masses (\( M_\text{CO} \lesssim 1\,M_\odot \)), consistent with the expectation that most binary systems have luminous, low-mass companions, while massive compact objects are rare. Some of the stars with inferred low-mass companions may not be in binary systems but instead exhibit spuriously high astrometric or spectroscopic uncertainties. 

With the goal of identifying rare systems with massive dark companions, we focussed on stars with inferred companion masses \( M_\text{CO} \geq 3\, \mathrm{M}_\odot \), corresponding approximately to the canonical minimum mass for stellar-mass BHs.
Of the 21,028 candidates, 1088 stars exceed this threshold. To assess the significance of these high-mass candidates, we calculated for each star the posterior probability $\mathcal{P}(M_\mathrm{CO} > 3\,\mathrm{M}_\odot | \vec{d})$, defined as the posterior weight above this threshold. We find 493 stars (45\%) with $\mathcal{P} > 0.68$, which corresponds to approximately $1\sigma$ confidence, and 180 stars (17\%) with $\mathcal{P} > 0.95$ ($2\sigma$).
These probabilities are prior-dependent and assume that each star hosts a single dark companion responsible for the observed RV and astrometric excess scatter. The quoted $\sigma$ levels refer to the significance relative to measurement noise and not the probability of contamination by other effects, which we discuss in Sects.~\ref{sec:cleaning} and \ref{sec:discussion-contamination}.

Among the 1088 RGB+BH candidates are the known \textit{Gaia} BH~2 and the LAMOST mass-gap candidate, which illustrates and affirms the effectiveness of our selection method. In contrast, \textit{Gaia} BH~3 is not recovered, as we find a formally much lower inferred companion mass ($M_\mathrm{CO} \approx 1.3\,\text{M}_\odot$). This should have been expected: the orbital period of the system ($P \sim 11.6$ yr; \citealt{Panuzzo+2024}) far exceeds the \textit{Gaia} DR3 observing baseline. Despite hosting a massive BH, the limited phase coverage results in modest values of \texttt{ruwe} and $\sigma_\mathrm{RV}$, which are also consistent with a lower-mass companion in a shorter orbit. Given the higher prior probability for luminous, low-mass companions -- implemented through a log-uniform prior on companion mass and period in the model -- such solutions are statistically preferred. This highlights the difficulty of identifying BH companions in very wide orbits with DR3 data alone (BH~3 could only be detected with pre-release DR4 data) and the need for an extended baseline with the upcoming \textit{Gaia} DR4. It also shows that our selection is most sensitive to systems with orbital periods comparable to or shorter than the DR3 baseline. We discuss completeness and period sensitivity in more detail in Sect.~\ref{sec:completeness}.

\subsection{Cleaning the high companion mass candidate catalogue}
\label{sec:cleaning}

Only one RGB+BH system (and one additional candidate) has been confirmed based on orbital solutions published in DR3. This suggests that even if our method increases the search volume by a factor of a few, the number of new BH discoveries is likely to be $\sim$a few among the much larger catalogue of candidates for high-mass companions. Consequently, most of the 1088 identified stars with high inferred companion masses are expected to be contaminants rather than binaries with true compact object companions. The structures seen in the distribution of Fig.~\ref{fig:hist_RGB_fit_parameters} probably reflect the properties of these contaminating populations rather than the underlying remnant population. Given the susceptibility of the observables to false positives, we applied a series of cleaning steps to improve the reliability of the candidate list. These steps, described in the following, aim to increase sample purity while minimising loss of completeness.\footnote{The cuts described below refer to the 1088 candidates, but quality flags are provided for the full RGB sample in the published catalogue.}

\begin{figure}
    \centering
    \includegraphics[width=0.9\columnwidth]{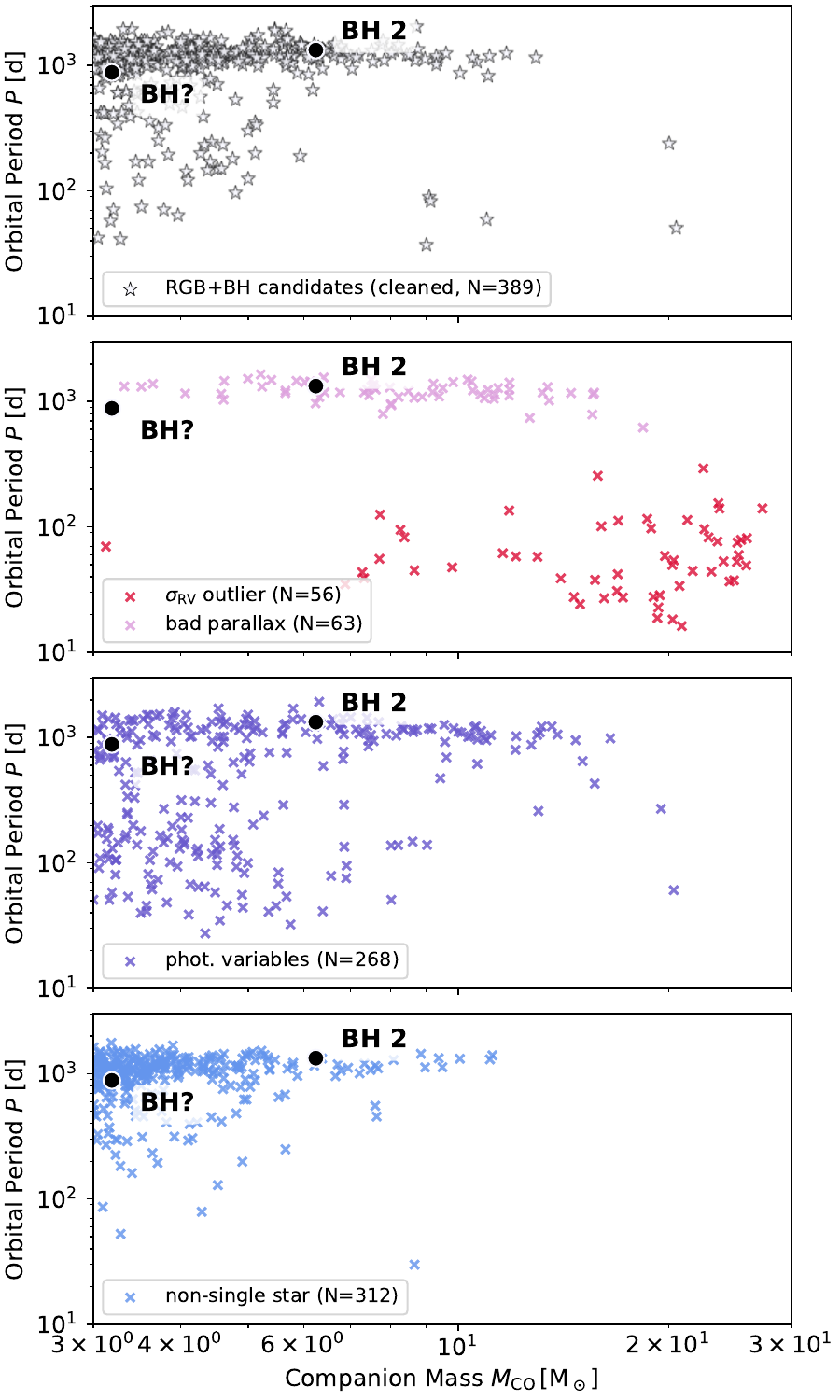}
    \caption{Inferred orbital periods and companion masses for the 1088 RGB stars with excess RV and astrometric scatter and companion mass estimates above $3\,\mathrm{M}_\odot$. The top panel shows the final cleaned catalogue of 389 RGB+BH candidates (see Sect.~\ref{sec:cleaning}), obtained after applying a series of quality cuts. The lower panels display the subsets of sources removed as likely contaminants by these cuts. The stars excluded because of spurious RV scatter (red) are concentrated at short inferred periods and high companion masses. Those with unreliable, underestimated parallaxes (pink) are typically found at long inferred periods. Eclipsing binaries and stars with photometric variability in excess of that expected from ellipsoidal modulation (purple) are scattered across both short and long periods. Systems excluded due to existing NSS orbital solutions (blue) occupy a broadly similar parameter space to the final candidates.} 
    \label{fig:cleaning}
\end{figure}

\begin{enumerate}
    \item[(A)] Spurious RV uncertainties --- Some stars exhibit large RV uncertainties that are not caused by binarity, leading to overestimated companion masses when these uncertainties are interpreted as arising from orbital motion. 
    To identify spurious cases, we compared the reported RV uncertainty with the \texttt{rv\_amplitude\_robust} parameter in the \textit{Gaia} DR3 catalogue, which represents the maximum RV difference observed for each source. We empirically inferred the expected relation between $\sigma_\text{RV}$ and \texttt{rv\_amplitude\_robust} using a population of simulated BH binaries (see Sect.~\ref{sec:completeness} for details). We forward modelled the expected RV amplitude using the same framework used to predict $\sigma_\text{RV}$, specifically, by computing the maximum amplitude of the modelled epoch RVs. The majority of RGB+BH candidates follow this relation. However, we removed 56 outliers that deviate by more than 3$\sigma$. These outliers typically have low values of \texttt{ruwe} but unusually high RV scatter, coupled with short inferred orbital periods and high companion masses (Fig.~\ref{fig:cleaning}, second panel). 

    \item[(B)] Unreliable parallaxes --- We removed 63 stars with potentially erroneous \textit{Gaia} DR3 parallax measurements. Since \texttt{ruwe} approximately scales with the angular size of the photocenter orbit, and hence inversely with parallax for a fixed physical orbit, underestimated parallaxes can lead to overestimated companion masses. To identify problematic cases, we compared the DR3 geometric parallaxes to the spectro-geometric parallaxes inferred from \textit{Gaia} XP spectra in the catalogue of \citet{Zhang+2023}, and excluded stars where the two estimates differed by more than a factor of two. These outliers typically show high inferred companion masses in wide, low-inclination orbits; a result of combining low RV scatter with elevated \texttt{ruwe} (Fig.~\ref{fig:cleaning}, second panel).
    
    \item[(C)] Photometric variables --- Photometric variability, particularly from stellar pulsations, can produce RV variations of several km\,s$^{-1}$ and thus mimic the signature of binary motion. In addition, eclipses in the light curve are incompatible with compact object companions. To mitigate contamination from such sources, we first cross-matched our RGB+BH candidate list with the \textit{Gaia} DR3 catalogue of variable stars \citep{Rimoldini+2023}. This allowed us to identify and remove 35 known eclipsing binaries, long-period variables, and other variables. We further excluded stars with excessive photometric variability by comparing their observed amplitudes with model predictions for ellipsoidal modulation. For each candidate, we calculated the expected photometric variability amplitude expected from ellipsoidal distortions, $A_\mathrm{phot}$, assuming an edge-on orbit (i.e. maximum modulation) based on the inferred companion mass and orbital period (see Sect.~\ref{sec:methods_photometry}). To define a conservative threshold, we added a noise floor of 0.127\,mag to the predicted amplitudes, corresponding to the 95.5th percentile of observed variability among the 21,028 RGB stars in the initial candidate sample. Stars with observed variability exceeding this combined threshold were excluded, resulting in the removal of 233 additional sources. Ellipsoidal modulation is only expected in short-period systems and becomes negligible for $R_*/a \ll 1$, so stars with large photometric amplitudes despite wide inferred orbits are conspicuous and likely spurious. These are effectively excluded by this cut (Fig.~\ref{fig:cleaning}, third panel).

    \item[(D)] NSS solutions --- We excluded all 312 candidates with existing solutions in the \textit{Gaia} DR3 NSS catalogue: 32 \texttt{Orbital}, 230 \texttt{SB1}, and 50 \texttt{AstroSpectroSB1} systems. Our focus is on identifying new candidates without prior orbital solutions, and NSS sources with high mass functions have already been studied elsewhere. Most of the excluded systems were marginal, with inferred companion masses just above the threshold value, reflecting the uncertainties in mass inference. NSS companion masses are generally $<3\,\mathrm{M}_\odot$, apart from the known BHs Gaia BH~2 and the LAMOST candidate and a handful of binaries with near face-on inclinations and correspondingly large uncertainties. 
    Discrepancies between our estimates and the NSS values may result from spurious \textit{Gaia} signals (elevated \texttt{ruwe}, RV scatter unrelated to binarity) or unresolved hierarchical systems, where the NSS solution captures only one component. This cut excludes the known \textit{Gaia} BH~2 and the LAMOST mass-gap BH candidate, both of which would otherwise meet our selection criteria.
\end{enumerate}

\begin{figure}[ht!]
    \centering
    \begin{minipage}[t]{0.9\columnwidth}
        \centering
        \includegraphics[width=\columnwidth]{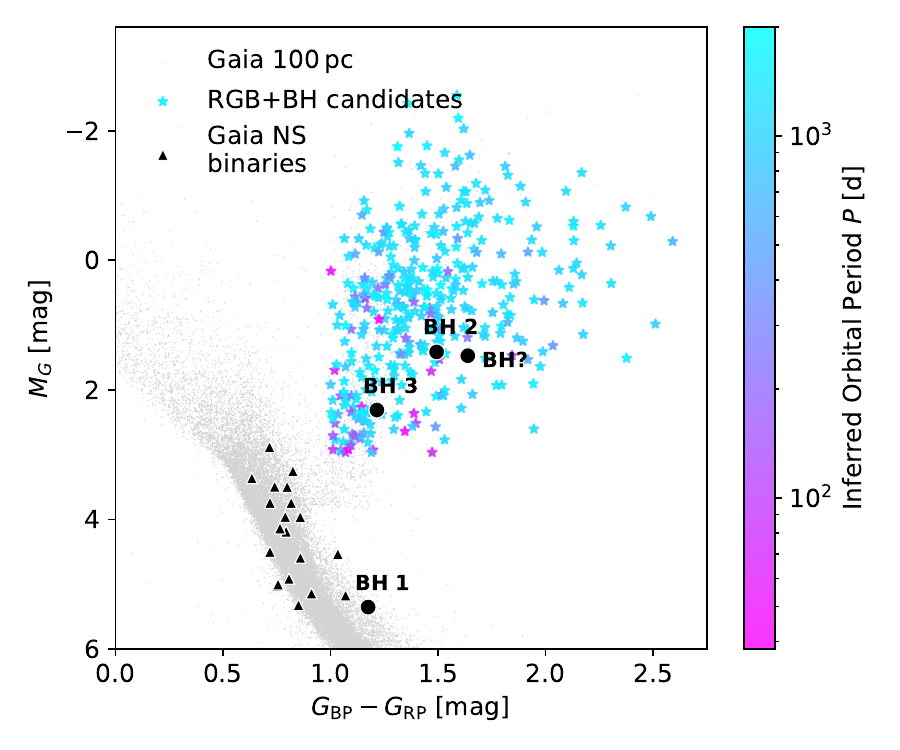}
    \end{minipage}
    \begin{minipage}[t]{0.9\columnwidth}
        \centering
        \includegraphics[width=\columnwidth]{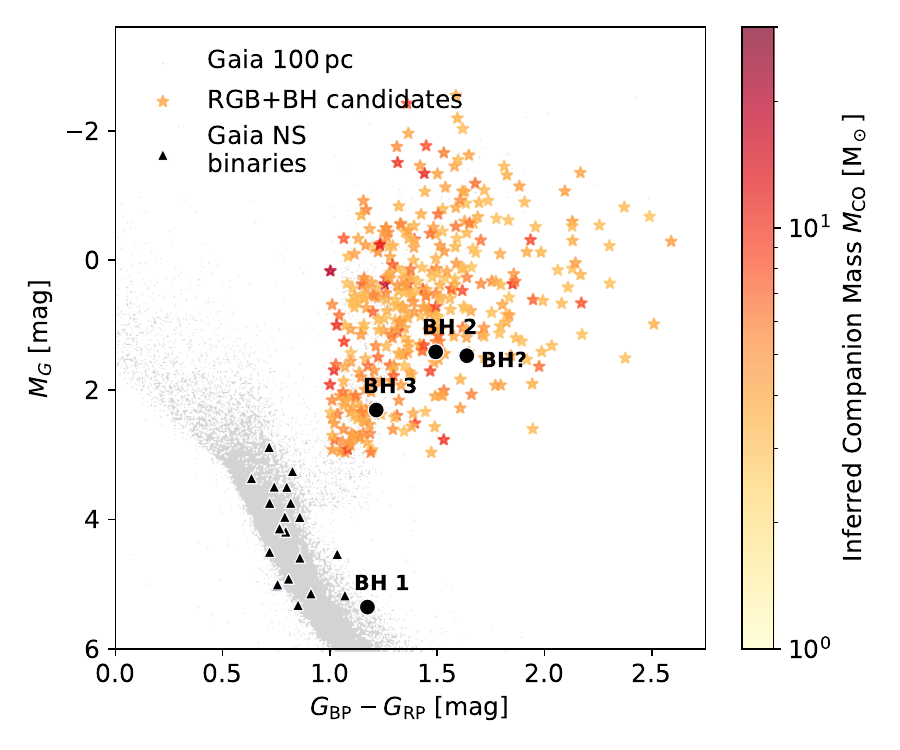}
    \end{minipage}
    \caption{CMD of the cleaned RGB+BH candidate sample of 389 stars (Sect.~\ref{sec:cleaning}). The 100\,pc bright star sample from \textit{Gaia} DR3 is shown for reference (grey points). Highlighted are the confirmed BH binaries \textit{Gaia} BHs 1, 2, and 3; the candidate BH from LAMOST (circles); and the \textit{Gaia} NS binaries (triangles). The candidates are coloured according to their inferred orbital periods (top) and companion masses (bottom). Short-period systems ($\lesssim 100$\,d) occur only on the lower RGB, consistent with expectations that the larger radii of more evolved giants preclude close orbits.} 
    \label{fig:cleaned_candidates_CMD}
\end{figure}

The cleaned sample consists of 389 stars; their locations in the CMD are shown in Fig.~\ref{fig:cleaned_candidates_CMD}. The candidates are broadly distributed across the RGB, with no significant clustering or trends between CMD position and inferred companion mass, which would point to unrecognised systematics in the parameter inference. 
Systems with short inferred orbital periods ($\lesssim 100$\,d) are located toward the lower RGB, consistent with expectations given the physical size constraints of more evolved giants. 
These 389 stars exhibit \textit{Gaia} astrometric and spectroscopic signatures along with minimal photometric variability consistent with massive dark companions, making them promising targets for dormant NS or BH binary searches. 

\subsection{Comparison with \textit{Gaia} orbital solutions}
\label{sec:nss_validation}

\begin{figure*}[ht!]
    \centering
    \begin{minipage}[t]{0.9\textwidth}
        \centering
        \includegraphics[width=\columnwidth]{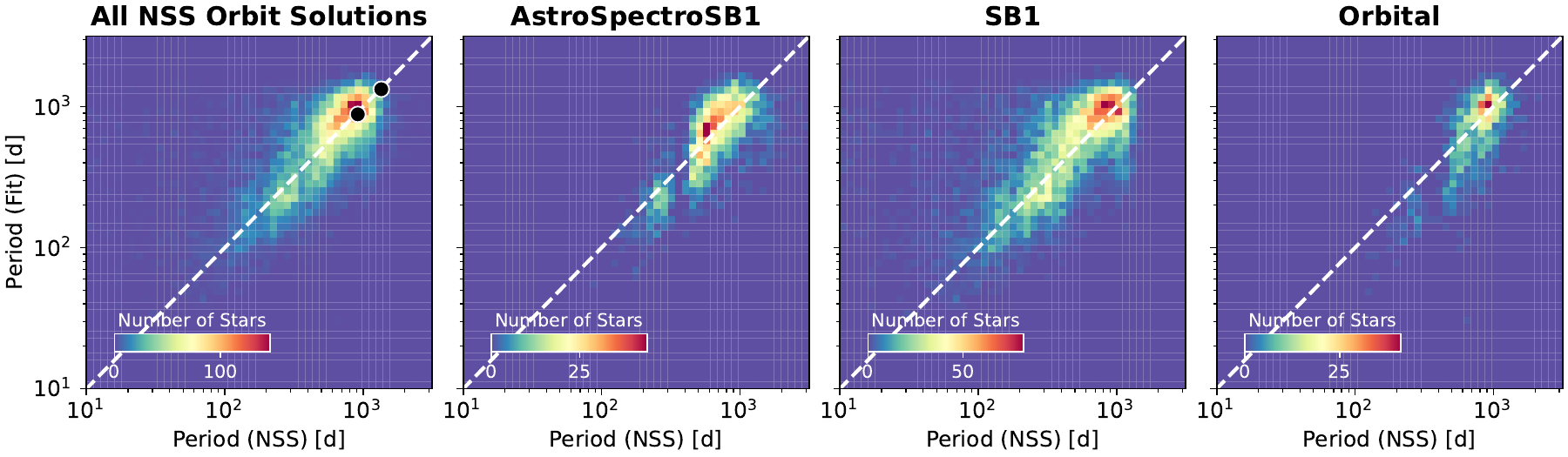}
    \end{minipage}
    \begin{minipage}[t]{0.9\textwidth}
        \centering
        \includegraphics[width=\columnwidth]{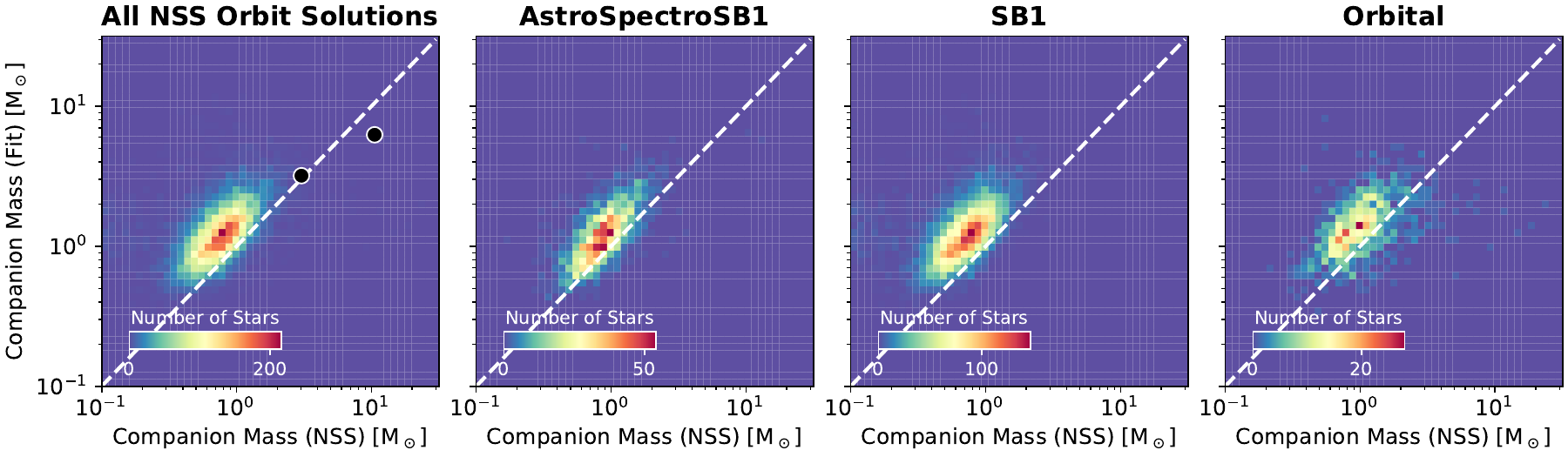}
    \end{minipage}
    \caption{Comparison between binary parameters inferred from \textit{Gaia} summary diagnostics and reference values from the DR3 NSS catalogue. Left panels: Histograms for the 10,420 stars (out of 21 thousand variable RGB stars) with \textit{Gaia} orbital solutions. Right panels: The subsets with \texttt{AstroSpectroSB1}, \texttt{SB1}, and \texttt{Orbital} solutions. Comparisons are shown for orbital periods (top) and companion masses (bottom). \textit{Gaia} BH~2 and the candidate BH from LAMOST are highlighted (black). The systematic offsets in the inferred parameters likely reflect a combination of inclination effects, the preferentially high inclinations and shorter periods of systems in the NSS sample, and the fact that spectroscopic orbits yield only minimum companion masses.} 
    \label{fig:nss_fit_comparison}
\end{figure*}

Although sources with published \textit{Gaia} DR3 NSS orbital solutions were excluded from our final catalogue of RGB+BH candidates (Sect.~\ref{sec:cleaning}), they provide an important consistency check for our inference framework. We therefore compared our inferred orbital parameters with the published NSS solutions for the subset of stars where both are available. Of the 21,028 RGB stars for which we inferred binary parameters, 10,841 have NSS binary solutions. After applying the cleaning steps from Sect.~\ref{sec:cleaning} (removing photometric variables, unreliable parallaxes, and $\sigma_\mathrm{RV}$ outliers), this number reduces to 10,420 stars with NSS solutions.\footnote{This subset includes the 312 high-companion-mass candidates with published NSS solutions that were removed from the final RGB+BH candidate catalogue in Sect.~\ref{sec:cleaning}.}
Figure~\ref{fig:nss_fit_comparison} (top) shows the comparison of catalogued vs inferred orbital periods for these 10,420 objects \citep[similar to][their Fig.~9]{Andrew+2022}. Despite the limited number of observables and many degrees of freedom in the inference problem, we find a good overall correspondence. Some scatter is expected, given the large uncertainties and typically broad posterior distributions (see Fig.~\ref{fig:comparison_fit}). At long periods ($\sim$1000\,d), our method tends to overestimate the orbital period. This probably reflects selection effects in the NSS catalogue: binaries with $P \gtrsim 1000$\,d are given more often astrometric acceleration or spectroscopic trend solutions rather than full orbital solutions, so the NSS comparison sample is biassed toward shorter periods.

We also compared the inferred companion masses with the NSS values (Fig.~\ref{fig:nss_fit_comparison}, bottom). Again, a linear relation is evident, though our method systematically yields higher masses by an average of 35\%. Several effects contribute here, mainly related to orbital inclination. First, NSS spectroscopic solutions lack inclination constraints, so their companion masses represent lower limits. Second, the sample of stars with spectroscopic solutions is biassed toward edge-on systems, where RV amplitudes are the highest. In contrast, in our forward modelling we assumed an isotropic distribution of inclinations (uniform in $\cos i$). This marginalisation overestimates masses for edge-on configurations; if these are more prevalent in the comparison sample, it causes a systematic offset. 
An additional contribution may come from our initial selection cut of \texttt{rv\_amplitude\_robust} $>20\,$km\,s$^{-1}$, which biases the sample toward high \texttt{rv\_amplitude\_robust} and $\sigma_\mathrm{RV}$ outliers. Since massive companions are intrinsically rare, this selection likely favours systems with overestimated amplitudes due to noise or sampling effects.
Finally, for some NSS systems, the uncertainties in the inclination also add scatter, meaning that the uncertainties on the $x$-axis can also be large.

\subsection{Catalogue completeness}
\label{sec:completeness}

\begin{figure*}[t]
    \centering
    \includegraphics[width=0.9\textwidth]{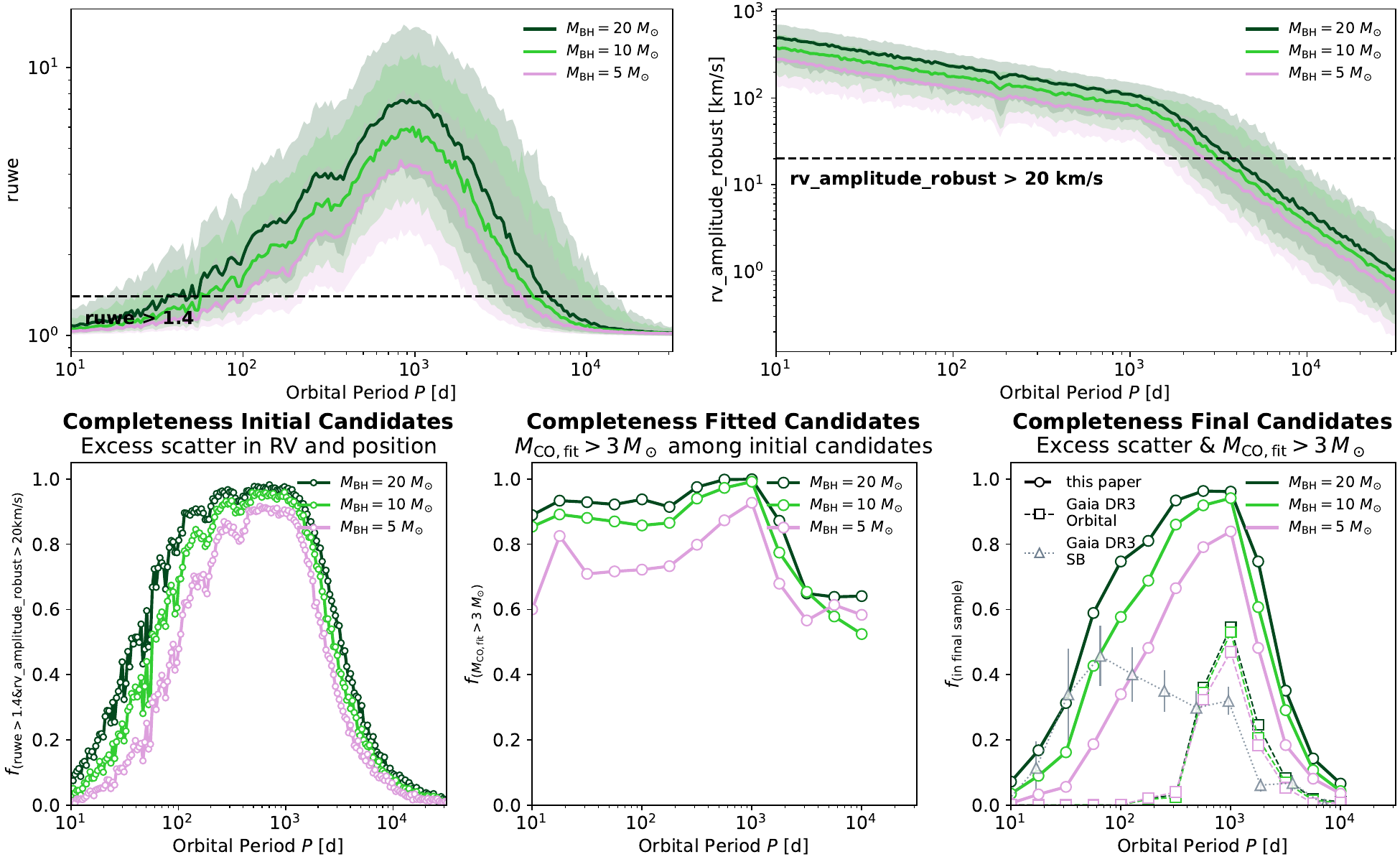}
    \caption{Simulated completeness of the RGB+BH binary catalogue as a function of BH mass and orbital period. Top panels: Mean values of predicted \texttt{ruwe} (left) and \texttt{rv\_amplitude\_robust} (right) from mock \textit{Gaia} DR3 observations, shown for simulated binaries with BH masses of 5, 10, and 20\,M$_\odot$. The shaded regions indicate one standard deviation from the mean. The threshold values for the cuts on excess scatter in RV and the position applied to the candidate sample are shown in black. Higher BH masses generally yield stronger astrometric and RV signals. The astrometric \texttt{ruwe} increases with orbital period up to the DR3 baseline ($\sim$1000\,d), while the RV amplitude is highest for short-period systems. 
    Bottom panels: Recovery fractions for simulated binaries that pass the initial cuts on excess scatter in RV and position (left), that exceed the inferred companion mass threshold of 3\,M$_\odot$ (centre), and that meet both criteria (right). The final completeness is highest ($\gtrsim$80\% for 10–20\,M$_\odot$ BH) for binaries with periods of 100–2000\,d. Completeness decreases for longer periods due to incomplete phase coverage within the DR3 observing baseline. For reference, we show approximate completeness fractions for the \textit{Gaia} DR3 astrometric (dashed) and spectroscopic (dotted) orbital solutions estimated based on the \texttt{gaiamock} forward model and SBX catalogue of spectroscopic binaries (see Sect.~\ref{sec:completeness} for details).}
    \label{fig:completeness}
\end{figure*}

We assessed the completeness of the presented catalogue of 389 RGB+BH candidates with a simulated population of BH binaries. Specifically, we estimated the recovery fraction of BH companions as a function of orbital period and BH mass by forward modelling the selection process on a sample of mock binaries. 

To this end, we simulated RGB+BH binaries with BH masses of 5, 10, and 20 M$_\odot$, and orbital periods uniformly spaced in $\log P$ between 10 and $3\times10^4$ days. Eccentricities were drawn from a thermal distribution and we assigned orbital angles with an isotropic distribution and uniformly distributed orbital phases. Each simulated binary was paired with a randomly selected RGB star from the observed \textit{Gaia} parent sample of 3.3 million RGB stars (see Sect.~\ref{sec:rgb_sample}), but we required reliable stellar parameters and $T_\mathrm{eff}<6000\,$K. The simulated binary inherited the observed stellar properties, brightness, and coordinates of the chosen RGB star.

For each combination of period and BH mass, we created $1000$ such binaries. We then used \texttt{gaiamock} to simulate the corresponding epoch astrometry, RVs, and DR3 summary diagnostics; \texttt{ruwe}, \texttt{rv\_amplitude\_robust}, $\sigma_\mathrm{RV}$, and $A_\mathrm{phot}$ (see Sect.~\ref{sec:forward_model}). These mock observables were processed with our inference framework (Sect.~\ref{sec:parameter_inference}) to estimate companion masses and orbital periods, with the goal of recovering the input parameters.

A RGB+BH binary enters the final catalogue only if it passes the two-stage selection function: (1) it must show excess scatter in RV and position; \texttt{ruwe}\,$>1.4$ and \texttt{rv\_amplitude\_robust}\,$>20$\,km\,s$^{-1}$, and (2) the inferred companion mass must exceed 3\,M$_\odot$. The resulting recovery fractions are shown in Fig.~\ref{fig:completeness}.

The upper panels show the mean values of \texttt{ruwe} and \texttt{rv\_amplitude\_robust} for the simulated binaries as a function of period and BH mass. At fixed period, higher BH masses result in larger orbits, and thus larger astrometric signals, reflected in increased \texttt{ruwe} values. As expected, \texttt{ruwe} increases with orbital period up to the \textit{Gaia} DR3 observation baseline of $\sim10^3$\,d, beyond which the phase coverage is less complete and \texttt{ruwe} decreases. A noticeable dip occurs at orbital periods close to one year, where orbital motion becomes degenerate with parallax. Systems with periods between $\sim$100 and several thousand days typically exceed the \texttt{ruwe}\,$>1.4$ threshold. These trends depend on the distance distribution of the RGB stars in our sample, which extends to several kiloparsecs.

The RV amplitude of a binary scales with companion mass and inversely with orbital period, explaining the downward trend in \texttt{rv\_amplitude\_robust} with period in the upper-right panel. Up to periods of $\sim$10 years, most systems fall above the 20\,km\,s$^{-1}$ threshold.

The lower panels of Fig.~\ref{fig:completeness} show the recovery fraction of simulated RGB+BH binaries across three stages; (1) passing the initial \texttt{ruwe} and \texttt{rv\_amplitude\_robust} cuts (left), (2) having inferred companion masses above 3\,M$_\odot$ (centre), and (3) satisfying both criteria and entering the final catalogue (right). Results are shown for the three considered BH masses as a function of the orbital period.

The completeness of the initial candidates (lower left) is high ($\gtrsim80\%$) for orbital periods of 100-2000 days. The completeness of the fitted candidates (lower centre), that is, the probability of inferring $M_\mathrm{CO} > 3\,M_\odot$ for binaries that show excess scatter in position and RV, increases with true BH mass and is roughly constant below $\sim$1000 days, but falls off at longer periods due to incomplete orbit coverage. In this regime, reduced astrometric and RV excess scatter can mimic lower-mass, shorter-period companions, which are favoured by the prior. The completeness of the final candidate sample (lower right) combines the previous two stages, that is, it shows the probability for a RGB+BH binary to pass both cuts. For a 20\,M$_\odot$ BH, completeness exceeds 80\% between 100 and 1000 days, while for a 5\,M$_\odot$ BH, more than half of the systems are still recovered in this range. The sharp decline at longer periods highlights the limitations of using DR3 summary statistics and explains why long-period systems such as \textit{Gaia} BH~3 are unlikely to be recovered using this approach.

A central motivation for our use of summary diagnostics to identify dormant BH candidates was to explore beyond the strict quality cuts applied to orbital solutions published in \textit{Gaia} DR3. Therefore, it is interesting to compare the completeness of our summary diagnostic-based selection with that of the NSS catalogue.

To do so, we used the \texttt{gaiamock} package, which models not only the epoch astrometry and summary statistics of \textit{Gaia} DR3 but also emulates the full astrometric processing pipeline, including the fitting of orbital solutions and the associated quality cuts. We passed the same population of mock RGB+BH binaries described above through the simulated \textit{Gaia} pipeline and determined the fraction of systems that would have received a full 12-parameter orbital solution in DR3. The resulting completeness curve is shown as the dashed line in the lower right panel of Fig.~\ref{fig:completeness}.

Similarly to our candidate selection method, the completeness of the \textit{Gaia} astrometric solutions peaks at orbital periods of of $\sim$1000 days, where the astrometric binary signature is the largest. For longer periods, completeness declines because of insufficient phase coverage and the resulting preference for acceleration (7- or 9-parameter) solutions over full orbits. Our summary diagnostics-based approach produces significantly higher overall completeness; approximately twice as high as astrometric solutions for periods $\gtrsim$500 days, and more than 20 times higher in shorter periods, where the recovery rate of orbital solutions in DR3 is close to zero.

Estimating the completeness of \textit{Gaia} spectroscopic binary orbits requires a different approach from the astrometric binaries. The orbital fitting process for spectroscopic binaries in DR3 is a complex multi-stage pipeline \citep{Gosset+2025}, and the associated selection function is not yet well characterised. In the absence of a dedicated emulator equivalent to \texttt{gaiamock} for the spectroscopic pipeline, we instead roughly estimated the spectroscopic completeness empirically using a reference catalogue of known binaries.

For such an approach, we used the \href{http://astro.ulb.ac.be/sbx}{SBX} catalogue of spectroscopic binaries \citep[previously called SB9;][]{Pourbaix+2004, Merle+2026}, which contains orbital parameters for more than 5000 systems. Assuming that the SBX catalogue is close to complete within its target parameter space, we estimated \textit{Gaia} spectroscopic completeness as the fraction of SBX binaries for which DR3 provides a spectroscopic orbital solution. To ensure a fair comparison with our RGB+BH candidate sample, we applied similar selection criteria to the SBX sample, restricting it to 936 luminous giants ($\mathtt{phot\_g\_mean\_mag} + 5 \log(\varpi / 1000) + 5 < 3$\,mag, $T_\mathrm{eff} < 6000$\,K) that are bright and have measured epoch RVs in DR3, necessary for an SB solution ($\mathtt{rv\_method\_used} = 1$). 

The period distribution of this filtered SBX sample peaks near 1000 days, and the distribution of RV semi-amplitudes at $\sim$10\,km\,s$^{-1}$. We did not filter on RV amplitude, as completeness does not show a simple monotonic relation with amplitude. The resulting spectroscopic completeness, defined as the fraction of SBX binaries with a DR3 spectroscopic orbit, is shown in the lower right panel of Fig.~\ref{fig:completeness} (dotted line). The completeness peaks at short orbital periods ($<100\,$ d). The general completeness in this sample is approximately 23\%, but exceeds this value in the range between one month and 100 days. However, we note that receiving an SB1 solution does not necessarily mean that the source will also be identified as a candidate BH. That is because approximately a third of these systems will be too face-on to unambiguously identify the companion as a BH. In this sense, the estimated spectroscopic completeness may be overestimated. 

Comparing the completeness curves for NSS spectroscopic and astrometric solutions with the completeness of summary diagnostic-based selections, 
the latter is most effective in extending the search volume for dormant BHs in the intermediate period range between approximately 100 days and two years. 
We propose systems in this period regime as the most promising targets for spectroscopic follow-up.

We note that the astrometric and spectroscopic completeness curves cannot be combined by simple addition. The two methods are not mutually exclusive, and many binaries with \texttt{AstroSpectroSB1} solutions appear in both subsets.

\section{Discussion}
\label{sec:discussion}

\subsection{Contamination of the final candidate sample}
\label{sec:discussion-contamination}
Several sources of contamination will affect our RGB+BH candidate sample. Here we summarise the most relevant effects (see Sect.~\ref{sec:cleaning} for details on the applied quality cuts).

Our inference is based on \textit{Gaia} DR3 summary statistics, particularly \texttt{ruwe} and $\sigma_\mathrm{RV}$. While both are sensitive to binarity, they can also be affected by other circumstances. For example, marginally resolved wide binaries can distort the astrometric solution and increase astrometric excess noise \citep{Arenou+2023,Halbwachs+2023,Holl+2023,El-Badry+2024c,Gosset+2025}. This effect motivated the stringent parallax signal-to-noise cuts applied in DR3 for astrometric binary solutions. Furthermore, stellar activity, such as pulsations or star spots, can inflate RV scatter. To mitigate contamination from these non-orbital effects, we excluded stars with variability inconsistent with ellipsoidal modulation, as well as obvious eclipsing binaries (Sect.~\ref{sec:cleaning} C).

Erroneous parallaxes may also cause sample contaminants, as the inferred companion mass depends on the assumed physical scale of the orbit. In our inference, the parallax was treated as a free parameter, so that the uncertainties in $\varpi$ were consistently propagated in the inferred orbital parameters and companion masses. To address extreme cases, we further removed stars with strongly discrepant \textit{Gaia} parallaxes relative to spectrophotometric estimates \citep[][Sect.~\ref{sec:cleaning} B]{Zhang+2023}. However, moderate parallax underestimates in the remaining sample may still lead to overestimated companion masses.

Companion masses can also be overestimated for binaries with nearly edge-on inclinations, for which the inferred companion mass is systematically inflated when marginalised over an assumed isotropic inclination distribution. This introduces contamination from RGB binaries with companions on the main sequence or with white dwarf companions (see Sect.~\ref{sec:nss_validation}). 

Luminous stellar companions can create false positives through several mechanisms. A giant with a warm companion may be mistaken for a higher-mass giant, since the star will appear bluer at similar luminosity, and overestimating the giant's mass leads to an overestimated companion mass.
A related source of contamination comes from (post)mass-transfer binaries with stripped giant stars. Recently, several proposed BH candidates around evolved stars were revealed to be stripped giants with masses significantly lower than inferred from single-star evolutionary models \citep[$\sim0.4\,$M$_\odot$, e.g. the Unicorn;][]{El-Badry+2022,Jayasinghe+2022}. In these systems, the giant undergoes stable mass transfer to a subgiant companion, making the more evolved star the less massive component. Although not BH or NS binaries, identifying such systems would provide valuable insights into a rarely observed phase of binary evolution leading up to the formation of helium white dwarf or low-mass subdwarf binaries.

Contamination from hierarchical triple systems is another major concern \citep{Andrew+2022}. In such systems, a short-period inner binary can enhance RV variability, while a distant tertiary induces astrometric motion, jointly mimicking the signal of a single massive companion. These configurations can lead to spurious inferences of an intermediate-period binary with a compact object. Similar contamination has been identified in earlier searches for dark companions based on high \texttt{ruwe} and RV variability \citep{Andrew+2022}. To minimise this, we restricted our analysis to evolved stars; while stars low on the RGB may still host short-period inner binaries, the large current or past radii of more evolved giants preclude such systems. However, triples with moderately separated components (e.g. 200-day and 2000-day orbits) could still remain in the sample.

Despite the cleaning steps applied, it is expected that a high fraction of contamination remains, requiring spectroscopic follow-up to identify and eliminate false positives more robustly (see Sect.~\ref{sec:follow-up}).

\subsection{Expected Gaia orbital solutions for RGB+BH candidates}
\label{sec:nss_prediction}
As a consistency check on the inferred binary parameters, we estimated how many of the RGB+BH candidates would be expected to receive orbital solutions in \textit{Gaia} DR3 and which fraction are likely to do so in the upcoming DR4 release. We followed the approach used in Sect.~\ref{sec:completeness} and forward modelled the \textit{Gaia} astrometric processing pipeline with \texttt{gaiamock}. For each of the 389 RGB+BH candidates in the cleaned catalogue, we generated 20 realisations of the binary parameters as random draws from the weighted posterior samples. Each realisation was then passed through the simulated \textit{Gaia} astrometric processing chain to estimate the probability of a given solution type (single-star 5-parameter, acceleration 7- or 9-parameter, or 12-parameter orbit).

Summing over all candidates, we find that $\sim$74.4\% are expected to receive single-star solutions in DR3, 5.9\% acceleration solutions, and only 19.8\% full 12-parameter orbital solutions. The moderate predicted fraction of orbital solutions is reassuring: if many more of our candidates should already have received 12-parameter solutions, this would cast doubt on the inferred parameters. For individual stars, the probability of an orbit solution is low, with only one candidate exceeding a 50\% probability.
Under the assumption that the inferred parameters are correct and using the same NSS quality cuts as for DR3, we predict a substantial increase in orbital solutions in DR4. The simulations suggest that about 63\% of our candidates could then receive astrometric binary orbits.

\subsection{Spectroscopic follow-up}
\label{sec:follow-up}
High-resolution, multi-epoch spectroscopic observations are essential to confirm the nature of the RGB+BH candidates and to identify (presumably abundant) contaminants. Luminous companions can be detected through double-lined (SB2) features and spectral disentangling techniques, allowing systems with hot secondaries or post-mass-transfer histories to be distinguished from genuine NS or BH companions. Hierarchical triples, one of the primary expected sources of contamination, can be revealed if the observed RV variability is inconsistent with the inferred astrometric motion, for example through short-period variations indicative of an inner binary.
RV curves also provide independent constraints on the orbital period, RV amplitude, and minimum companion mass. The spectra also help refine the stellar parameters and infer the abundances of the primary stars. 

We have begun an observing campaign for spectroscopic follow-up of the most promising candidates in our sample using the FEROS instrument at the La Silla MPG 2.2\,m telescope with approximately 50 nights of allocated observing time. Although complete phase coverage may be infeasible because of the typically long orbital periods (on the order of several years), even few-epoch RV data can be useful when combined with the epoch astrometry and RVs expected from \textit{Gaia} DR4. These data will extend the observing baseline and enable joint orbital fits, improving parameter constraints and the prospects of confirming dormant NS or BH companions. The results of the follow-up will be described in a forthcoming paper.

\section{Summary and conclusions}
\label{sec:conclusion}
Most known stellar-mass BHs have been detected in X-ray binaries or through gravitational wave events. These techniques probe only a minute and biased subset of the total population. Detecting larger numbers of non-accreting dormant BHs will enable the exploration of the bulk properties of the Galactic BH population and help improve our understanding of binary evolution processes and supernova physics.

The \textit{Gaia} mission provides a great opportunity to identify such systems. While the DR3 NSS catalogue has dramatically increased the binary census and led to the discovery of the first Galactic dormant BHs in binaries, published orbital solutions are limited by strict quality criteria, thereby sampling only a very limited portion of parameter space. Significant improvement is expected with DR4, which will include epoch astrometry and RVs over an extended baseline with sensitivity to longer orbital periods and a larger set of orbital solutions. Meanwhile, \textit{Gaia} DR3 already offers powerful diagnostics for identifying binaries with massive companions, even without explicit orbital solutions. Summary diagnostics such as elevated astrometric scatter (quantified by \texttt{ruwe}) and large RV variations (reflected in $\sigma_\mathrm{RV}$) indicate orbital motion from unseen companions.

We developed a forward-modelling framework to infer binary parameters directly from these summary statistics. Given assumed binary properties, we predicted corresponding \textit{Gaia} observables and inverted the process to infer posterior distributions for companion mass and orbital period. We applied this method to bright RGB stars with DR3 single-epoch RV measurements, constructing a catalogue of 389 RGB stars whose variability is consistent with hosting massive dark companions -- potentially dormant BHs or NSs -- in orbits with periods of $\sim 10^{1.7\, -\, 3.3}$ days. Focussing on RGB stars offers the advantage that their large radii preclude close ($P \sim 10$ day) inner binaries, reducing contamination from hierarchical triples. Candidate selection was most complete for orbital periods between roughly 100 and 1000 days, where the combined astrometric and spectroscopic sensitivity of \textit{Gaia} is highest, and successfully recovered known dormant RGB+BH binaries in this regime. Compared to the NSS catalogue, our method probed roughly twice the search volume of astrometric orbits across all periods (and more than three times larger for $P<500$ or $P>2000$\,d), and double to triple the volume of spectroscopic orbits for $P \gtrsim 200$\,d. 

Spectroscopic follow-up remains essential to validate candidates and eliminate false positives. High-resolution, multi-epoch RV monitoring can confirm the binary origin of variability, identify spurious systems with luminous secondaries or hierarchical triples, and provide independent constraints on orbital parameters and minimum companion mass. Combining follow-up RVs with future \textit{Gaia} DR4 epoch astrometry will enable joint orbital fits and secure identification of dormant NS and BH companions.

\section*{Data availability}
The full catalogue of 21,028 RGB stars with excess RV and astrometric scatter, including inferred orbital periods and companion masses, is available on \href{https://doi.org/10.5281/zenodo.17271785}{https://doi.org/10.5281/zenodo.17271785}. The dataset also contains additional parameters and quality flags that allow users to select the 389 most promising RGB+BH candidates from the cleaned catalogue. An analogous catalogue for 19,664 main-sequence star candidates is also provided and described in Appendix~\ref{app:ms_candidates}.

\begin{acknowledgements}
The authors sincerely thank the anonymous referee for their thorough review and thoughtful and constructive comments. J. M\"uller-Horn thanks Thibault Merle for helpful discussions and suggestions regarding the use of the SBX binary catalogue. 
J. M\"uller-Horn, B. Pennell, M. Green, J. Li, R. Seeburger, and H. W. Rix acknowledge support from the European Research Council for the ERC Advanced Grant [101054731]. 
This work has made use of data from the European Space Agency (ESA) mission \textit{Gaia} (https://www.cosmos.esa.int/\textit{Gaia}), processed by the \textit{Gaia} Data Processing and Analysis Consortium (DPAC, https://www.cosmos.esa.int/web/\textit{Gaia}/dpac/consortium). Funding for the DPAC has been provided by national institutions, in particular the institutions participating in the \textit{Gaia} Multilateral Agreement.

\end{acknowledgements}

%
   \bibliographystyle{aa} 
   \bibliography{references.bib} 
%

\begin{appendix}

\section{Stellar parameter estimation with \texttt{XGBoost}}
\label{app:xgboost}

\begin{figure}
    \centering
    \begin{minipage}[t]{\columnwidth}
        \centering
        \includegraphics[width=\columnwidth]{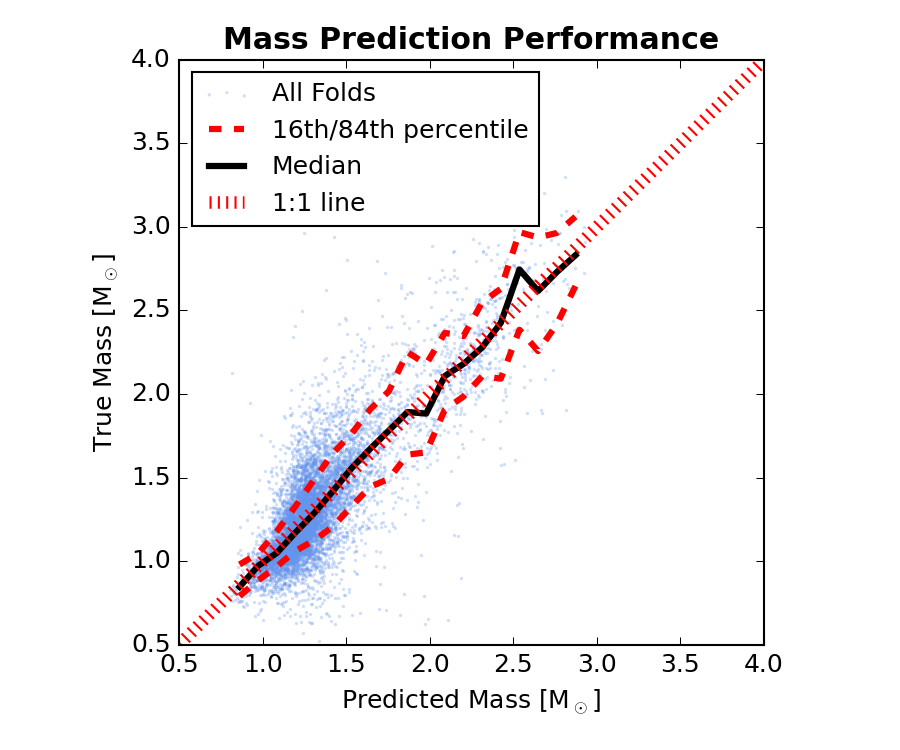}
    \end{minipage}
    \begin{minipage}[t]{\columnwidth}
        \centering
        \includegraphics[width=\columnwidth]{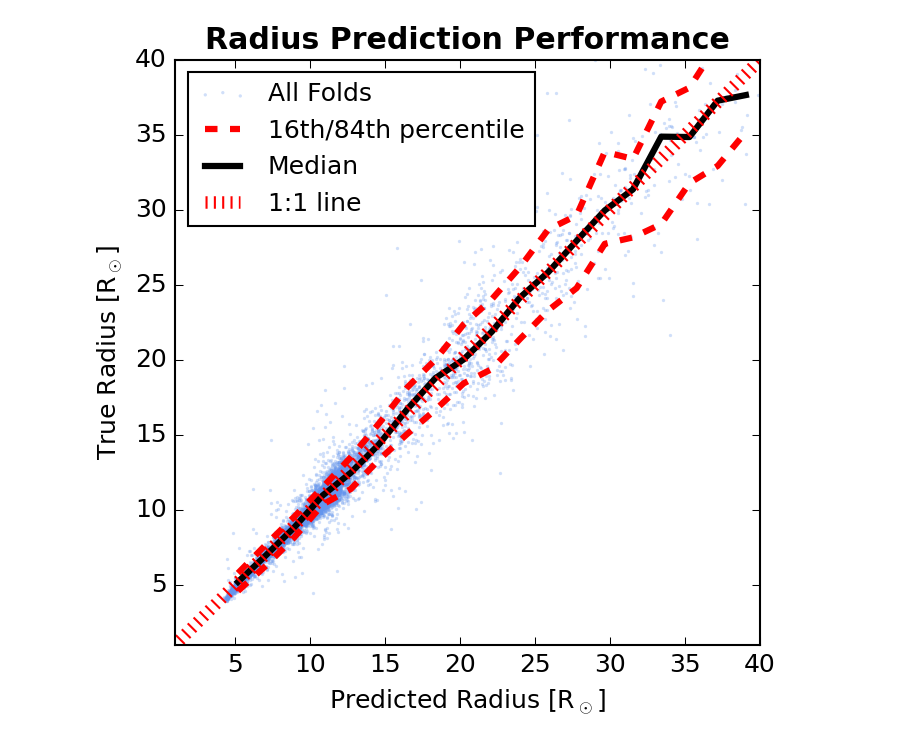}
    \end{minipage}
    \caption{Performance of the \texttt{XGBoost} model in predicting stellar masses and radii for RGB stars. The top and bottom panels show the true vs predicted values for mass and radius, respectively, based on the validation set. The reference values are asteroseismic measurements from the APOKASC-3 catalogue. The model achieves typical fractional uncertainties of $\sim$15\% for mass and $\sim$7\% for radius, with small systematic bias.} 
    \label{fig:xgboost}
\end{figure}

We trained an \texttt{XGBoost} regression model to predict stellar masses and radii for candidate \textit{Gaia} evolved stars using spectroscopic and photometric features. The input parameters included metallicity, effective temperature, and surface gravity from the \textit{Gaia} XP catalogue \citep{Zhang+2023}, along with 2MASS and WISE photometry (J, H, K, W1, W2) and colour indices (J–K, W1–W2). We used asteroseismic masses and radii from the APOKASC-3 catalogue \citep{Pinsonneault+2025} as target values and we restricted the training set to $\sim$10,000 RGB stars with $<10\%$ uncertainties in both mass and radius.

Model performance was assessed using 5-fold cross-validation. Within each fold, the input features were standardised (zero mean, unit variance), and an \texttt{XGBoost} regressor was trained with a maximum depth of six, a learning rate of 0.01, and 1000 estimators with early stopping. Performance was evaluated using the root mean square error (RMSE).

The model achieved mean cross-validation RMSEs of $0.21 \pm 0.01\,\text{M}_\odot$ for stellar mass and $1.23 \pm 0.15\,\text{M}_\odot$ for stellar radius. The predicted vs true values for the test set are shown in Fig.~\ref{fig:xgboost}. For stellar masses, we determined a mean and standard deviation of the logarithmic ratio $\log(M_\mathrm{pred}/M_\mathrm{true}) = 0.011 \pm 0.150$, corresponding to a typical fractional uncertainty of $\sim$15\% and minimal bias. For stellar radii, we obtained $\log(R_\mathrm{pred}/R_\mathrm{true}) = 0.005 \pm 0.069$, or $\sim$7\% uncertainty. As can be expected, the effective temperature and the absolute magnitude (in W1) were the most important training features.\footnote{These absolute magnitudes were computed without applying the \textit{Gaia} DR3 parallax zero-point correction \citep{Lindegren+2021b}, which would slightly reduce the inferred stellar masses of some stars. The effect is generally small compared to the \texttt{XGBoost} fit uncertainties, being less than 0.14\,M$_\odot$ for more than 99\% of RGB candidates.}

\section{A catalogue of dormant BH candidates on the main sequence}
\label{app:ms_candidates}

While our primary focus is on RGB stars, which typically dominate the light centroid and have reliable RV scatter, making compact-object companions easier to identify, our method is equally applicable to MS primaries. Here we present a catalogue of MS+BH candidates analogous to that for RGB stars.

The main challenge for MS stars is the increased contamination from luminous binaries and triples \citep{Andrew+2022}. To mitigate this, we used the data-driven MS+MS binary catalogue of \cite{Li+2025}, based on composite XP spectral signatures, to exclude systems with clearly luminous companions.

Our initial MS sample was drawn from the \textit{Gaia} DR3 archive using the following ADQL query, yielding 105,581 stars.
\begin{lstlisting}
SELECT *
FROM gaiadr3.gaia_source
WHERE parallax > 1
AND phot_bp_mean_mag < 20
AND bp_rp BETWEEN 0 AND 5
AND parallax_over_error > 10
AND phot_g_mean_mag + 5*log10(parallax/100) > 3
AND has_xp_continuous = 'true'
AND rv_method_used = 1
AND ipd_frac_multi_peak = 0
AND rv_amplitude_robust > 20
AND ruwe > 1.4
\end{lstlisting}

The first six cuts select MS stars with $M_G > 3$\,mag for which the XP filtering of luminous binaries is applicable \citep{Li+2025}, while the last four cuts match those applied to the RGB sample (Sect.~\ref{sec:rgb_sample}), selecting systems with excess scatter in RV and astrometry. 

\begin{figure}
    \centering
    \includegraphics[width=\columnwidth]{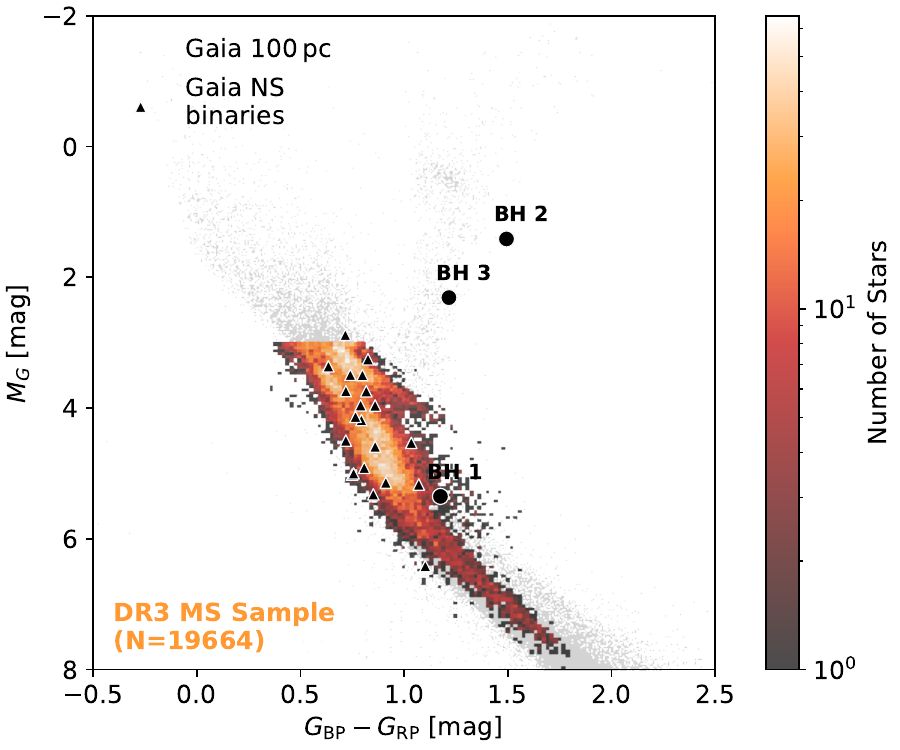}
    \caption{CMD of the initial MS candidate sample (2D histogram), overlaid on the \textit{Gaia} DR3 100\,pc reference catalogue (grey). Confirmed BH binaries \textit{Gaia} BH~1, BH~2, and BH~3 are highlighted, along with dormant NS binaries (triangles).} 
    \label{fig:CMD_MS_sample}
\end{figure}

We further required robust XP-based stellar parameters \citep[\texttt{quality\_flags} $< 8$;][]{Zhang+2023}, excluding 17,153 stars, and restricted the sample to objects with \texttt{predicted\_class} $=0$ in the \citet{Li+2025} MS+MS catalogue. This removed many luminous binaries and reduced the candidate sample to 19,664 stars. A CMD of this sample is shown in Fig.~\ref{fig:CMD_MS_sample}. Stellar masses for the primaries were taken from single-star isochrone fits in the MS+MS catalogue.

\begin{figure}
    \centering
    \includegraphics[width=\columnwidth]{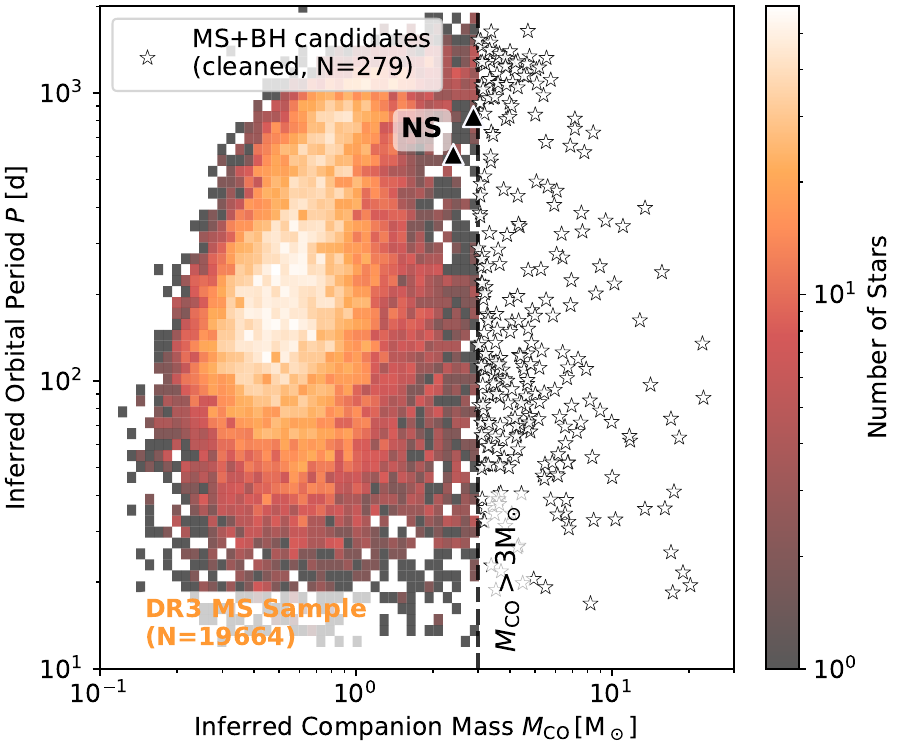}
    \caption{Inferred orbital periods and companion masses for the initial MS candidate sample of 19,664 stars with excess RV and astrometric scatter. The subset of promising high-mass candidates ($M_\mathrm{CO} > 3\, \mathrm{M}_\odot$) is highlighted with white markers on the right, separated by the dashed vertical line. Stars that survive the quality cuts described in Appendix~\ref{app:ms_candidates} and are not in the NSS catalogue yield a final cleaned sample of 279 MS+BH candidates. Also shown are the recovered parameters for the two known \textit{Gaia} MS+NS binaries in the sample, for which our method recovers moderate to high companion masses and long orbital periods.} 
    \label{fig:fit_MS_sample}
\end{figure}

\begin{figure}[h!]
    \centering
    \includegraphics[width=0.83\columnwidth]{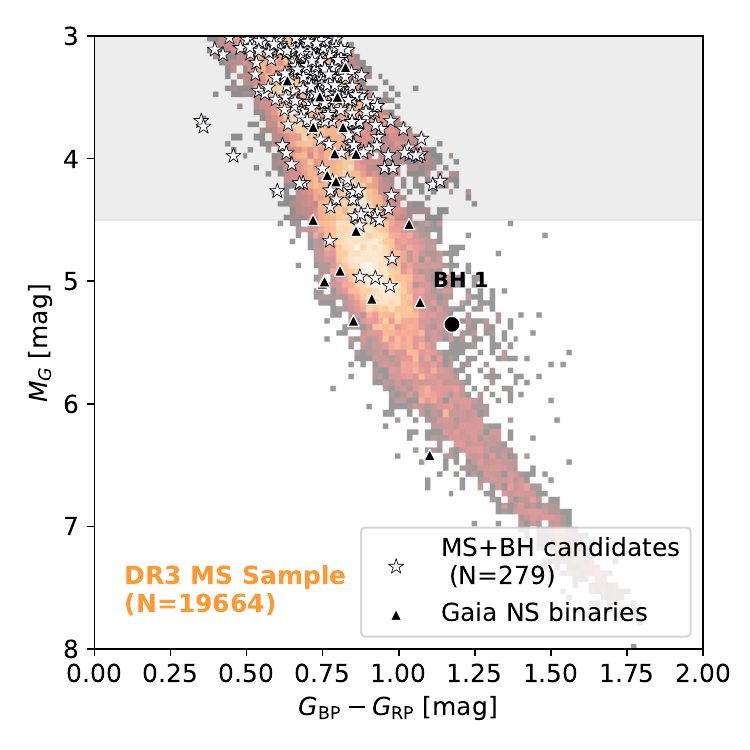}
    \caption{CMD of the 279 final MS+BH candidates (white stars), overlaid on the initial MS candidate sample (2D histogram). The confirmed BH binary \textit{Gaia} BH~1 is highlighted, together with known \textit{Gaia} NS binaries (triangles). The apparent clustering of candidates at $M_G < 4.5$\,mag (shaded grey) likely reflects the less efficient removal of MS+MS binaries on the upper main sequence by the XP-based filtering.} 
    \label{fig:CMD_MS_candidates}
\end{figure}

We then applied the same forward-modelling inference as for the RGB sample, fitting companion masses and orbital periods for all 19,664 MS stars. For these stars, we kept the parallax fixed to the DR3 catalogue value, since the sample is limited to nearby stars with high parallax significance ($\varpi/\sigma_\varpi > 10$) and the Gaia zero-point correction is less relevant at these distances. Figure~\ref{fig:fit_MS_sample} shows the inferred distributions. A total of 711 systems have companion masses above 3\,M$_\odot$. After applying the quality cuts described in Sect.~\ref{sec:cleaning}, that is, removing 225 stars with spurious RV scatter, one star with an unreliable parallax, 163 photometrically variable stars, and 43 with NSS orbital solutions, the final cleaned catalogue contains 279 MS+BH candidates. Their CMD distribution is shown in Fig.~\ref{fig:CMD_MS_candidates}. The apparent clustering of candidates at $M_G \lesssim 4.5$\,mag probably reflects residual contamination from luminous binaries. For bright, upper-MS stars, age effects complicate isochrone fitting, and XP spectral decomposition becomes less robust \citep{Li+2025}. In contrast, the small number of promising candidates at fainter magnitudes (only seven candidates with $M_G \geq 4.5$\,mag) makes spectroscopic follow-up both valuable and feasible in this regime.

Of the known compact-object binaries with MS primaries (including \textit{Gaia} BH~1 and several MS+NS systems), only two were included in our parent sample of 19,664 stars. The others were excluded because they lack epoch RV measurements (\texttt{rv\_method\_used} $\neq 1$) and therefore did not meet the initial selection criteria. For the two MS+NS systems that were included \citep[\textit{Gaia} DR3 5136025521527939072 and 1801110822095134848, with periods of 537 and 894 days, respectively;][]{El-Badry+2024b}, our fits recovered long periods ($P_\mathrm{fit} \approx 600$ and 800\,d) and high companion masses ($M_\mathrm{CO,fit} \approx 2.4$ and 2.9\,M$_\odot$). As expected for NS companions, these fall below the $3\,M_\odot$ BH threshold, but would be retained if the threshold were lowered slightly.

We also cross-matched our MS sample with the catalogue of compact-object binaries from \cite{Andrew+2022}, which contains 4,641 candidates. Applying the same cuts to this sample (e.g. on parallax, absolute magnitude, high \texttt{ruwe}, \texttt{rv\_amplitude}) yielded 1,341 stars in common with our initial MS sample. However, most of the \citeauthor{Andrew+2022} candidates were subsequently removed by the XP-based filtering of luminous MS binaries. Within our parent sample of 19,664 stars, only 178 overlapped with \citeauthor{Andrew+2022} (15, 56, and 107 in their gold, silver, and bronze categories). The inferred companion masses and orbital periods show very good agreement. After applying the full quality cuts described in Sect.~\ref{sec:cleaning}, only 30 stars remained in common with our final cleaned MS+BH candidate catalogue.

\begin{figure}
    \centering
    \includegraphics[width=0.66\columnwidth]{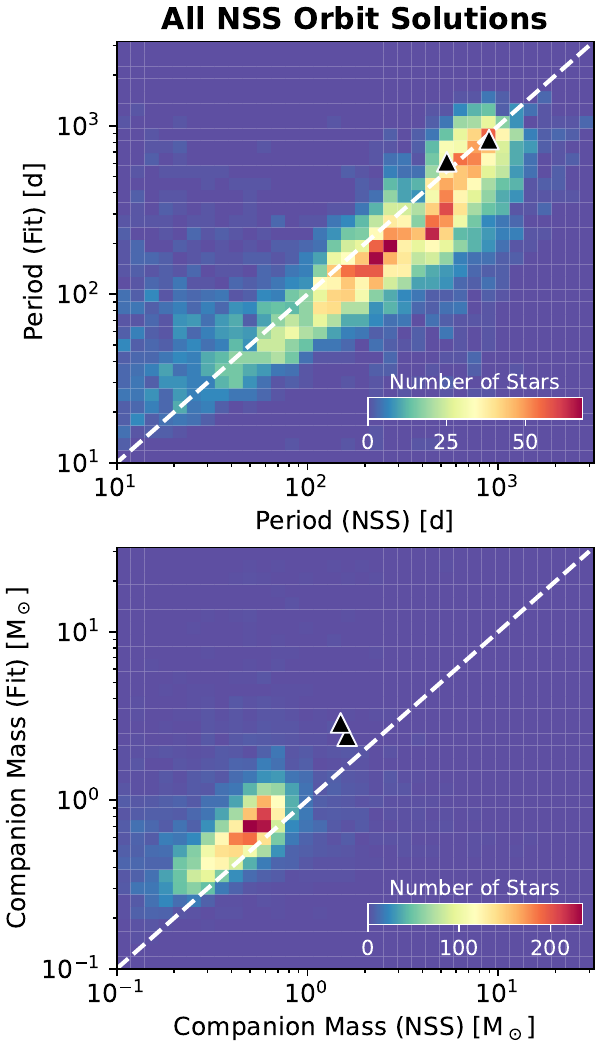}
    \caption{Comparison between binary parameters inferred from \textit{Gaia} summary diagnostics and reference values from the DR3 NSS catalogue for the MS sample. Shown are histograms for the 6195 stars (out of 19,664 fitted MS stars) with \texttt{AstroSpectroSB1}, \texttt{SB1}, or \texttt{Orbital} NSS solutions. The panels compare orbital periods (top) and companion masses (bottom). The two known \textit{Gaia} NS binaries in this sample are highlighted in black. Although they rely on only two observables (\texttt{ruwe} and $\sigma_\mathrm{RV}$), our fits reproduce the catalogued parameters with good overall correspondence.} 
    \label{fig:MS_nss_comparison}
\end{figure}

To further validate the method, we compared our inferred periods and masses to \textit{Gaia} DR3 NSS solutions for the 10,420 MS stars in the parent sample with \texttt{AstroSpectroSB1}, \texttt{SB1}, or \texttt{Orbital} entries. Figure~\ref{fig:MS_nss_comparison} shows that our fits reproduce the NSS parameters with good correspondence, despite being based only on two observables (\texttt{ruwe} and $\sigma_\mathrm{RV}$). However, for very short periods ($\lesssim 50$\,d), we often overpredict the orbital period. This bias likely arises from contamination by hierarchical systems, where short-period inner binaries inflate the RV scatter while long-period tertiaries increase \texttt{ruwe} values, leading our inference framework to favour intermediate periods.

The complete MS candidate catalogue with inferred parameters and quality flags is available on \href{https://doi.org/10.5281/zenodo.17271785}{Zenodo}.

\end{appendix}
\end{document}